\documentclass{article}

\usepackage{arxiv}

\usepackage[utf8]{inputenc} 
\usepackage[T1]{fontenc}    
\usepackage{hyperref}       
\usepackage{url}            
\usepackage{booktabs}       
\usepackage{amsfonts}       
\usepackage{nicefrac}       
\usepackage{microtype}      
\usepackage{lipsum}		
\usepackage{graphicx}
\usepackage{doi}

\usepackage[mincitenames=1,maxcitenames=3,
            minbibnames=1, maxbibnames=3,
            style=authoryear, uniquename=false,
            backend=biber]{biblatex}
\title{Seismology in the Solar System}

\author{ 
    \href{https://orcid.org/0000-0002-0783-2489}{
        \includegraphics[scale=0.06]{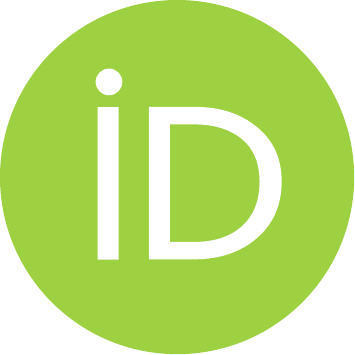}\hspace{1mm}Simon C. St\"ahler} \\
    Institute for Geophysics\\
	ETH Zürich
	\And
	\href{https://orcid.org/0000-0003-0319-2514}{
	    \includegraphics[scale=0.06]{orcid.pdf}\hspace{1mm}Martin Knapmeyer} \\
	Institute for Planetary Science\\
	DLR Berlin
	}
	
\renewcommand{\shorttitle}{Seismology in the Solar System}
\newcommand{\citep}{\parencite}
\newcommand{\citet}{\textcite}
\hypersetup{
pdftitle={Seismology in the Solar System},
pdfsubject={astro-ph.EP},
pdfauthor={Simon C.~St\"ahler AND Martin Knapmeyer},
pdfkeywords={First keyword, Second keyword, More},
}

\newcommand{\deltaV}{$\Delta V$}

\begin{document}
\maketitle

\begin{abstract}
	Where could we do seismology on other planets and why should we do it?
\end{abstract}

\section{Introduction}
Seismology is the classic method to investigate the deep interior of the Earth, as well as its dynamic behaviour. Seismic observations confirmed the existence of Earth's core \citep{oldham_constitution_1906}, gave the first indication of a mineralogically distinct mantle \citep{mohorovicic_beben_1909} as well as layering of the mantle and core \citep{dahm_new_1934, lehmann_p_1936, bullen_seismology_1956}. For a more detailed overview, see chapter 3 of this book \citep{knapmeyer_planetary_2022}. Seismometers were therefore among the first instruments installed on the surface of the Moon by the Apollo astronauts in 1969 \citep{latham_moonquakes_1971, toksoz_lunar_1972} and were part of the Viking instrument suite on Mars 1976 \citep{anderson_seismology_1977}. For a variety of reasons, among them the apparent failure of the Viking seismic experiment \citep{lazarewicz_viking_1981}, the focus on human spaceflight in the 1980s, the absence of landers until 1995 and a focus on Martian geochemistry in the 2000s and 2010s, no seismic measurements were done on a planet for the 4 decades after. On 14 November 2014, the Philae lander on the comet 67/P Churyumov–Gerasimenko for the first time since Apollo measured elastic waves on a celestial object, excited by its sampling mechanism. On 5 May 2018, an Atlas V rocket finally launched the NASA InSight mission towards Mars, where it landed on 26 November 2018 and installed a seismometer on the surface in the weeks thereafter. This mission has since repeated many of the successes of a century of seismology within a good 3 years \citep{banerdt_initial_2020, giardini_seismicity_2020, lognonne_constraints_2020, knapmeyer-endrun_thickness_2021, khan_imaging_2021, stahler_seismic_2021-1, hobiger_shallow_2021}, constraining the Martian interior structure from near surface to core. 

The successful execution of the InSight mission has renewed interest in seismic measurements as a natural part of landed missions to other planets. This article reviews possible scientific goals of seismic measurements on the major bodies of the solar system. In it, it follows \citet{metzger_moons_2022} in defining a "planet" as an object of significant geological complexity. Large moons, i.e. objects of a few 100 km radius with tectonic processes shaping the surface and a complicated interior thermal budget are still planets, even if they happen to orbit another planet instead of the sun. 
For each object, we also summarize what little is known about its seismic sources and which missions are possible given current technical limitations, as well as the programmatic landscape of the Voyage 2050 program of the European Space Agency (ESA) \citep{tacconi_esa_2021} and the Decadal Survey for Planetary Sciences and Astrobiology 2023-2033 by the National Academies of Sciences of the USA \citep{decadal_2023}. The Voyage 2050 program and the Decadal survey are officially only recommendations to the individual agencies (ESA and NASA respectively), due to the participation of members from the global scientific community, both represent a true international effort and are therefore likely to shape programmatic efforts by all space agencies.
\begin{figure}
    \centering
    \includegraphics[width=\textwidth]{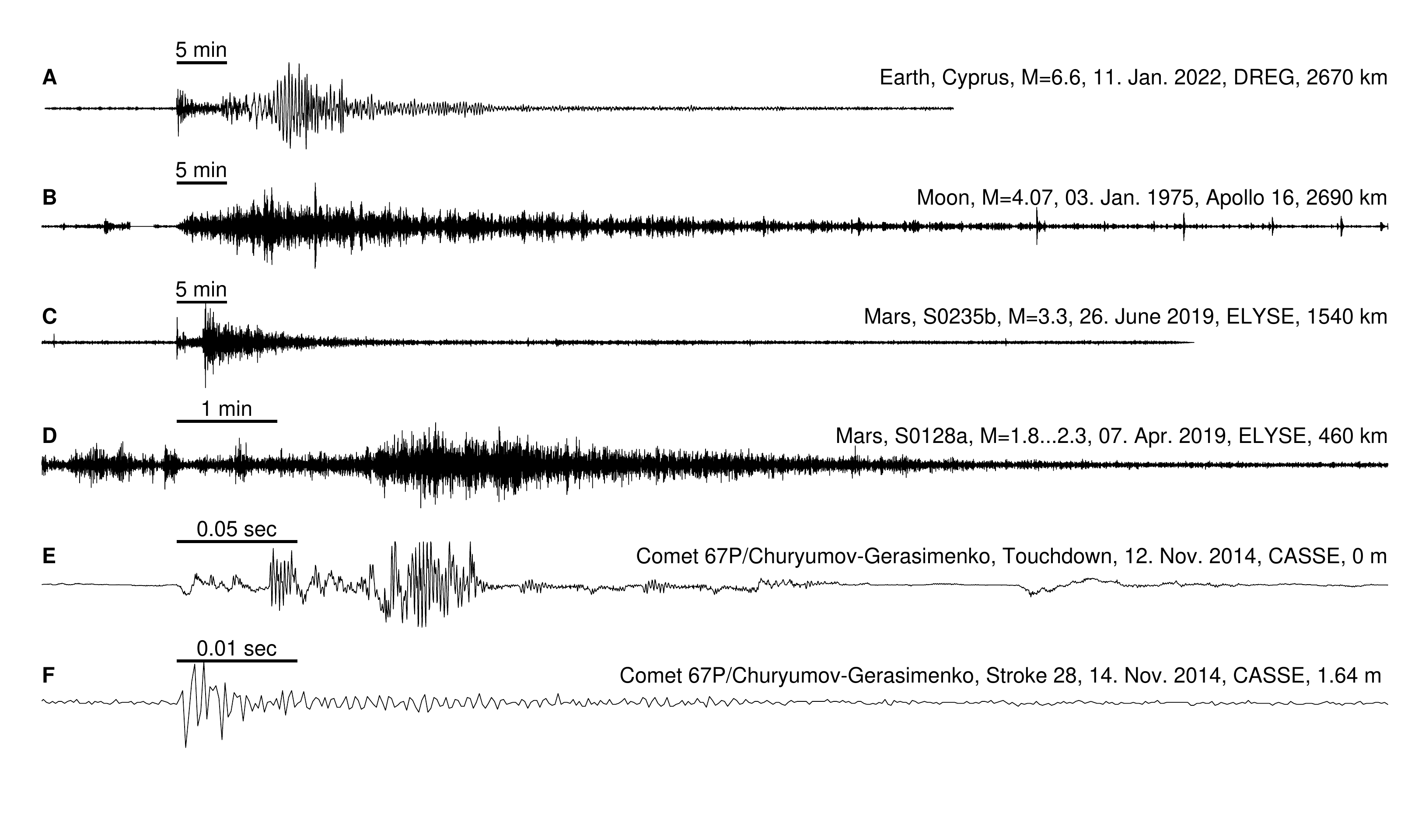}
    \caption{Seismograms recorded on four celestial bodies: Earth, the Moon, Mars and comet 67P.}
    \label{fig:seismograms}
\end{figure}
The article will hopefully serve as an introduction into the future of extraterrestrial seismology for seismologists, but even more, it should highlight to planetary scientists in general, which gaps in our understanding seismological data can fill. It is written to be readable by an interested reader with a general geological background.

\section{Mercury}
\begin{figure}
    \centering
    \includegraphics[width=0.7\textwidth]{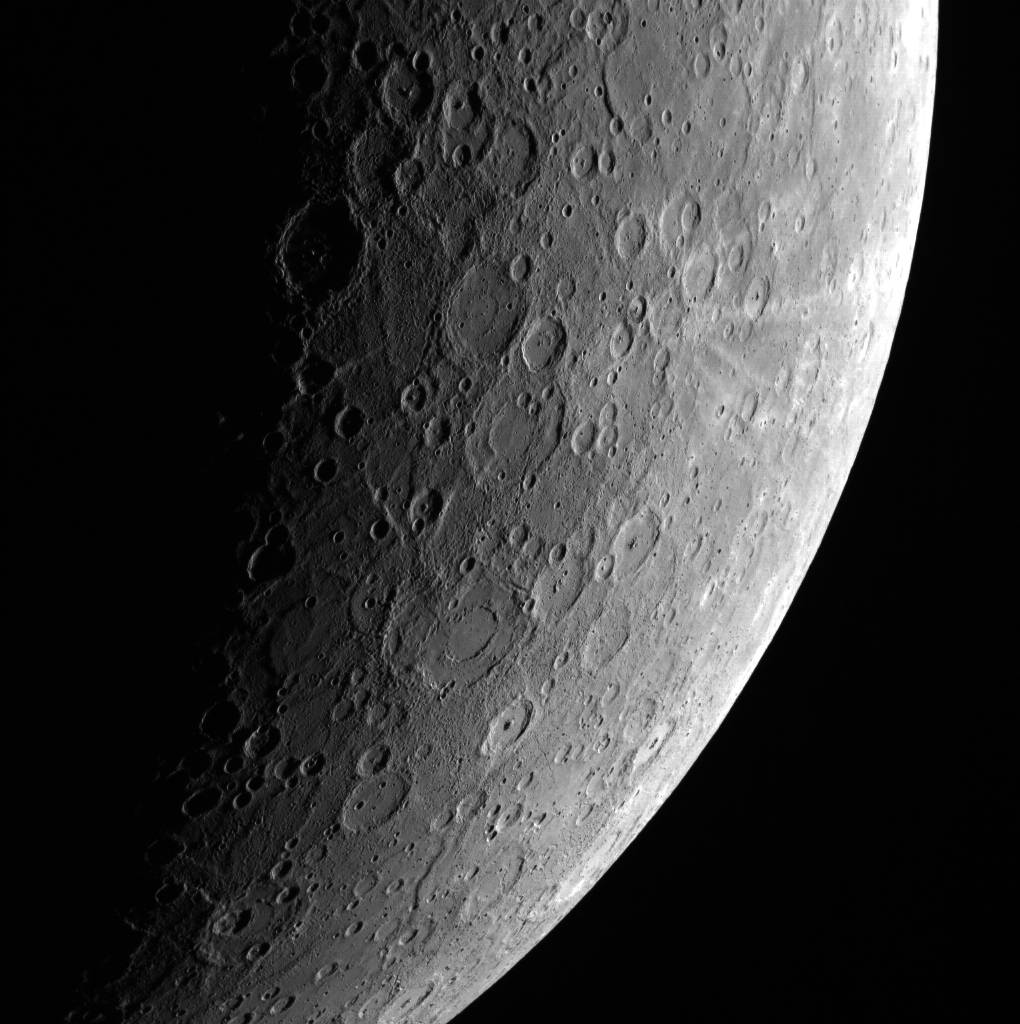}
    \caption{Mercury limb as seen by MESSENGER. The strong cratering of the surface is immediately visible and suggests that the megaregolith and its strong crustal scattering will be about as problematic for any seismometer mission, as it was on the Moon. The interior of Mercury with its large core might be much more interesting, though. Image: NASA/Johns Hopkins University Applied Physics Laboratory/Carnegie Institution of Washington PIA 17280}
    \label{fig:my_label}
\end{figure}
\begin{figure}
    \centering
    \includegraphics[width=\textwidth]{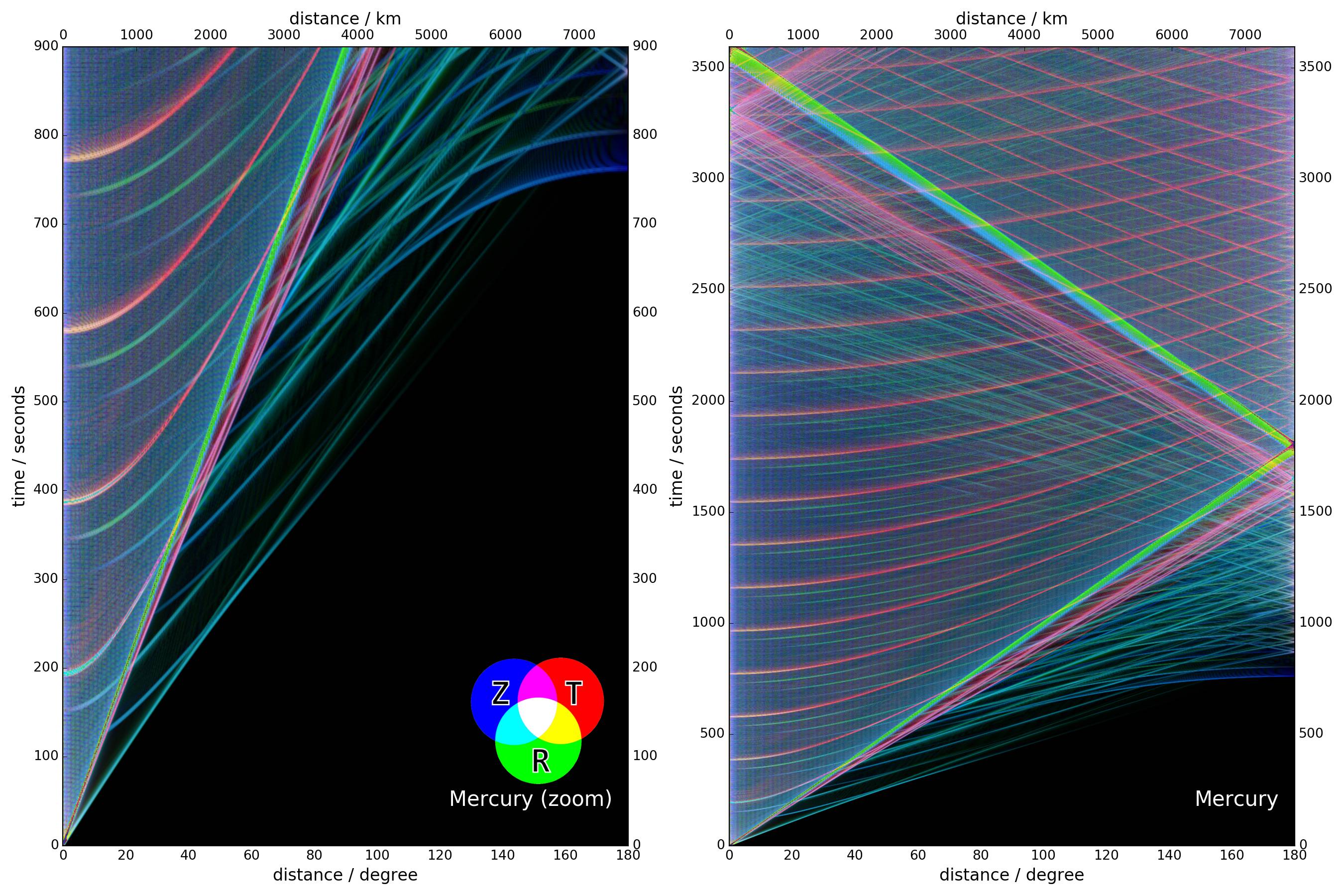}
    \caption{Global seismic wavefield stack for Mercury, using the interior model of \citet{rivoldini_interior_2009}. This stack shows surface acceleration in all 3 directions: blue for vertical, red for radial (horizontal in direction of the source) and red for transversal (horizontal, orthogonal to the source). The plot is a synthetic version of the global seismogram stack of \citet{astiz_global_1996} and computed using AxiSEM and Instaseis \citep{nissenmeyer_axisem_2014, van_driel_instaseis_2015}.
    Due to the large core size, the first core-reflected shear wave (ScS) arrives already after 200 seconds and is followed by regular multiples. A seismometer on a lander could use these phases to determine the core radius. Note that these numerical simulations do not reproduce scattering, so real waveforms could be much less clear.}
    \label{fig:gs_mercury}
\end{figure}
\subsection{Potential scientific goals}
Mercury is widely understood as a planet that was stripped of much of its mantle since formation \citep{chau_forming_2018}. The core radius is 85\% of the planet's radius with a mantle of only 400 km \citep{spohn_interior_2001} above. While the density is well-estimated from geodetic measurements \citep{rivoldini_interior_2009}, the distribution of weight between mantle and core is not. As shown on Mars \cite{stahler_seismic_2021-1, khan_geophysical_2018}, probing the mass and radius of a planet's layers very precisely, can lead to strong constraints on the composition of the planet. 
\subsection{Seismicity}
The level of background seismicity on Mercury is unknown. Compared to Mars and Earth, recent volcanism plays a minor role in shaping the surface of the planet \citep[see][for an extended comparison]{byrne_comparison_2020}. Painted with very broad strokes, seismicity on Earth and Mars is connected to volcanism: On Earth, the mid ocean ridges are expressions of extension driven at least partially by volcanism. Subduction zones on the other hand harbour the largest earthquakes known to date, at the thrust front, while the subduction process itself produces back-arc volcanism. Only transform faults are not producing co-located volcanism on their own, even though many strike-slip faults are located at the end of convergent faults.
On Mars, the strongest clusters of seismicity are found connected to recent volcanism. Dike intrusion is weakening the crust in Western Elysium Planitia, leading to a large number of marsquakes observed in Cerberus Fossae \citep{giardini_seismicity_2020, clinton_marsquake_2021}. 
Given the absence of plate tectonics or recent volcanism on Mercury, it is not clear what the driver of seismicity is. Wrinkle ridges, i.e. buried thrust faults are distributed over the whole planet and are interpreted as the result of global crustal contraction due to secular cooling \citep{byrne_mercurys_2014}. While the cumulative amount of deformation from this process can be estimated easily, it is not known whether the process is still ongoing or whether it has stalled. On the terrestrial moon, seismicity is triggered by tidal deformation, although it is disputed whether the tidal stresses directly cause the quakes or whether they just weaken normal stress on the fault temporarily so that rupture is possible. However, as shown by \citep[, for a summarysee table \ref{tab:tidal}]{hurford_seismicity_2020}, the tidal energy deposited in Mercury is significantly less that that of the moon.
However, the temperature difference of more than 600~K between day and night, is likely to form transient thermally-induced seismicity along the terminator.
    
\subsection{Mission perspectives}
Mercury places tight constraints on any landed mission. The place deep in the gravity well of the sun means that a lander mission will require at least one flyby on Venus or Earth to reduce its \deltaV to an acceptable level. The Ames trajectory database \citep{foster_mission_2010} lists no trajectories with a single Venus flyby and \deltaV $<20$ km/s until 2040. Therefore, any mission would involve rather complex and long duration trajectories. The ESA BepiColombo mission needs to do a total of 7 flybys (Earth once, twice at Venus and then six times Mercury itself) between 2021 and 2025 to enter an orbit.

After landing, the spacecraft would be subject to two separate realms of operations. Due to the 3:2 spin–orbit resonance of the planet, one solar day takes 176 Earth days. Therefore, temperatures vary extremely over time across the surface, between 100~K and 700~K in equatorial planes. Due to the low inclination of the orbit, temperatures at the poles, where the sun never rises more than 2 arcminutes above horizon, are more stable below 150~K. Since the low temperatures of the night can be relatively easily accommodated by electric heaters, most proposed missions target landing after sunset and limit the mission duration to a single night. \citet{ernst_mercury_2021} present a mission concept to study the geology around an equatorial landing site over the course of a single Mercury night (88 Earth days) in 2045. The mission study includes an accelerometer for seismic studies. The mission is proposed as a New Frontiers class mission and is currently the only serious lander concept for the innermost planet. The decadal survey \citep{decadal_2023} questioned whether this mission can be done within a New Frontiers budget and estimated a total cost of 2.8 billion USD, i.e. a flagship mission. Because of the narrower scientific scope of the mission as proposed compared to an ice giant system mission, it was ranked behind the Uranus Orbiter and the Enceladus Orbilander.

\section{Venus}
\begin{figure}
    \centering
    \includegraphics[width=\textwidth]{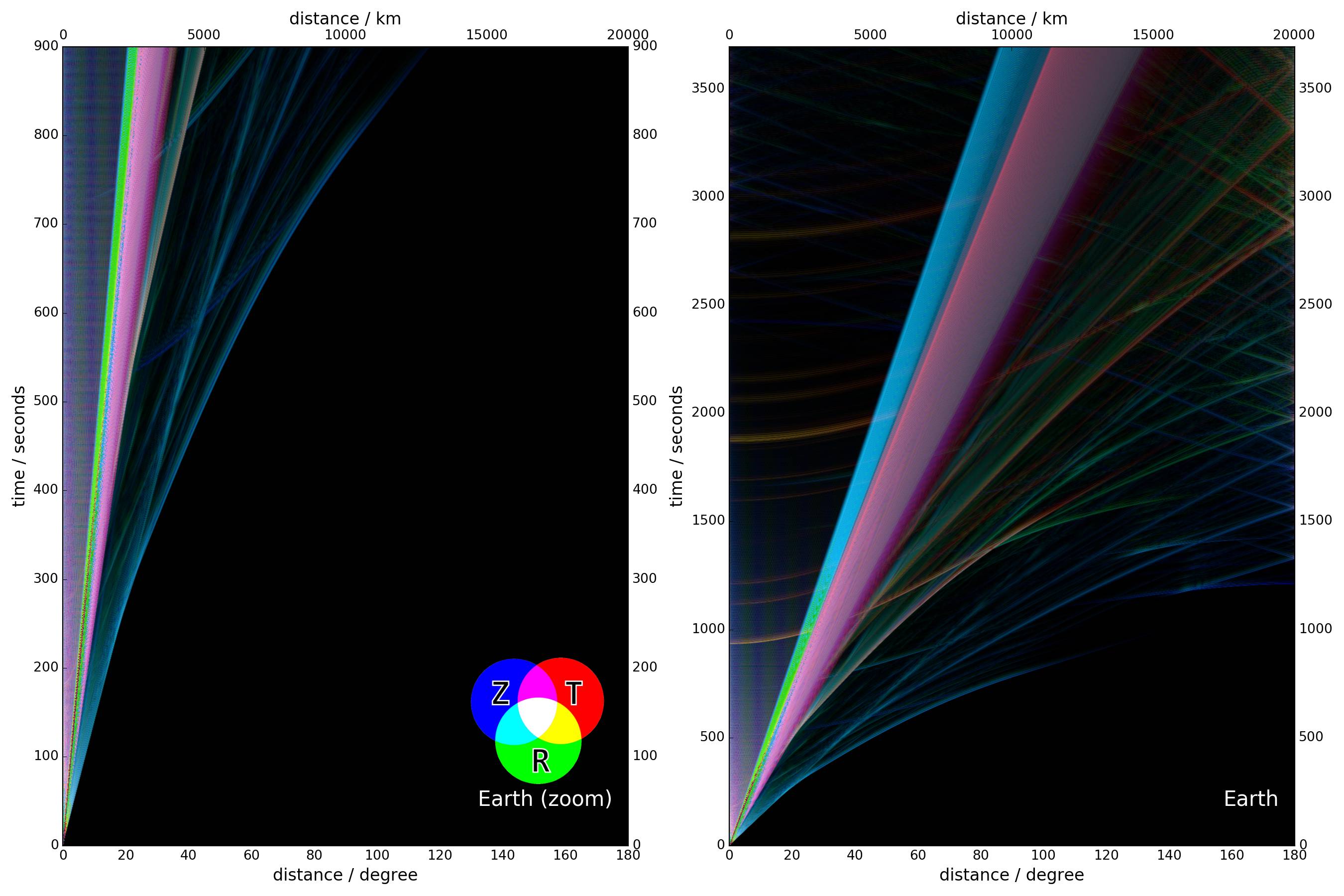}
    \caption{Global seismic wavefield stack for Earth, using the interior model of \citet{kennett_constraints_1995}. Venus may show similar seismic phases, given its comparable radius and density.}
    \label{fig:gs_earth}
\end{figure}
\subsection{Potential scientific goals}
Venus is comparable in size and density to Earth, yet little is known about its surface tectonics compared to Mars or even Mercury, due to the thick cloud cover. The surface has undergone significant reworking and its oldest parts, the so-called tesserae are likely younger than 600 million years. Based on available radar images from Magellan and the Arecibo radio telescope, the surface is undergoing strong deformation until today (SOURCE). In the next decade, the Envision \citep{ghail_science_2020} and Veritas \citep{smrekar_veritas_2020} missions planned by ESA and NASA respectively will increase the resolution of radar images significantly and help to constrain the dominant mechanisms of surface tectonics based on geomorphology. Plate tectonics in the terrestrial sense of the word is unlikely to exist on Venus, since the high temperature inhibit a strong lithosphere. Locations of venusquakes could however constrain whether active faults cluster, as they do on Earth and Mars or whether deformation is widespread, as on the Moon.

\subsection{Seismicity}
The seismicity of Venus is unknown. The relatively low crater density suggests an average surface age of 250–750 Myr, which can either be accomplished by a steady state of crater formation and removal by volcanic or tectonic processes \citep{phillips_is_1992} or catastrophic short term subduction or overturn of the whole lithosphere at regular intervals \citep{strom_global_1994}. Both scenarios suggest high, though different strain rates and therefore tectonic seismicity. However, given the high surface temperature, it is actually not clear how much of this deformation is occurring in a brittle regime and thus capable of producing venusquakes. The rate of volcanic eruption is likely higher than on Earth, with a recent study based on scaling the terrestrial rate to Venus estimating 120 discrete eruptions per (terrestrial) year \citep{byrne_estimates_2022}, which could be observed either seismically or from infrasound (see below).

\subsection{Mission perspectives}
Venus' orbit is relatively well accessible by spacecraft from Earth. Every 18 months, a launch window of 3 months opens with flight duration between 80 and 120 days. Landing and operating on the surface is of course an entirely different story. At this time, there are no seismic instruments at high technological readiness level that would be able to operate over extended periods under Venus conditions. The Soviet Venera landers relied on significant overdesign in terms of mass, which allowed survival of a few hours, by delaying the warming of the core electronics. The currently most promising candidate are Silicate Carbide (SiC) electronics that are able to operate at temperatures of 800~K, at least theoretically. The main application for SiC at this stage are high voltage, low loss systems, specifically power circuits, as well as simple amplification systems for sensors in high temperature environments \citep{zetterling_integrated_2015}. The complexity of the electronic systems that can be manufactured in SiC is far lower than for Si systems, and only basic integrated circuits have become available recently. A decisive problem is that the business case for high temperature SiC electronics is weak on Earth and development of a Venus lander would not be able to profit from commercial innovation cycles as much as it is the case for Si-based electronics. Specifically, industry development focuses on single components that cannot be moved to a cooler part of a system (such as sensors or pre-amplifiers), while a Venus lander would need to operate virtually all of its electronics at temperatures $>700~$K. Even with SiC electronics, a lander interior would have to be cooled actively. Solar irradiation is 3-5 W/m$^2$, so that a mission would have to rely on a radio-thermal generator, which would operate at low efficiency, given the high outside temperature. A European concept for a long-lived lander with a seismic package based on microelectronics was presented by \citet{wilson_venus_2016}, with a life time of 100 days, banking on further progress in SiC electronics over the next decade.

An alternative approach is to operate from the upper atmosphere, where temperatures are stable and moderate around 290~K. The lifetime of such a mission would be limited primarily by the escape of Helium from the carrying balloon, to the order of 120 (Earth) days, but at much lower technical complexity compared to a surface mission. A balloon mission equipped with infrasound sensors could detect air-coupled Rayleigh waves. Terrestrial analog concepts are currently studied on two scales: A JPL-led team studies the lower range of signal amplitudes that can be observed, using earthquakes of magnitudes 3-4 in California \citet{brissaud_first_2021}, as well as induced seismicity in Oklahoma. A French group centered on ISAE supaero uses free floating, long-lived meteorological balloons on equatorial trajectories to observe the global signature of large earthquakes above magnitude 7, as well as volcanic explosions \citep{podglajen_stratospheric_2022}. In both cases, signal detection was demonstrated, for a variety of events.
The obvious drawback of the method would be that the polarization of the actual seismic wave would not be accessible directly. \citet{garcia_active_2021} demonstrated two possibilities to overcome this limit: A string of infrasound sensors on the balloon tether could determine the incident angle of the infrasound wave, as a proxy of the distance to the event's hypocenter. A second option is the inclusion of accelerometers into the balloon, which would record the acceleration of the balloon due to the arrival of the pressure wave. Similar techniques are used in marine seismic surveys, usually termed multisensor streamers \citep{robertsson_use_2008}. The advantage is that the sensor effectively measures the pressure gradient of the infrasound wave, which allows recording the signal at higher frequencies (since the spatial derivative of observing the gradient is equivalent to recording the time derivative of the signal). The final application will likely see a combination of multiple infrasound sensors and accelerometers, from which the full pressure gradient field is reconstructed and transferred to Earth.

An unknown issue is that the coupling into the air is most effective for Rayleigh waves, preferably at frequencies above 0.1~Hz. As the InSight example showed, surface waves are not regularly observed, if hypocenters are not very shallow and event magnitudes are below $M_W=4$. 
A curious question would be where Venus stands in terms of seismic scattering. The dense atmosphere reduces the meteorite impact rate and thus impact gardening that produces the lunar regolith. The high surface temperature and availability of volatiles likely allows for healing of cracks and thus increases the mean free path length of seismic waves. So in terms of seismic waves, Venus might be the most "transparent" planet in the solar system, even compared to Earth.

\section{Moon}
\subsection{Potential scientific goals}
The moon hosted the first extraterrestrial seismic network between 1970 and 1979, when it was switched off due to lack of further interest in the seismic community \citep{lognonne_planetary_2007}. The Apollo seismic network helped exploring the Moon in a number of ways, from the shallow subsurface \citep{sollberger_shallow_2016} to the crust \citep{khan_new_2000} and the core \citep{weber_seismic_2011}. See \citep{khan_lunar_2013, garcia_lunar_2019} for an overview of the knowledge on the lunar interior gained from Apollo. The network further observed significant numbers of tectonic quakes and impacts \citep{kawamura_evaluation_2017}. Since a network was used, a tentative identification of shallow tectonic quakes with surface faults was possible in a few cases, allowing to estimate the energy budget of the moon due to shrinking \citep{watters_shallow_2019}. 

Seismometers have been part of all proposals for future network science on the moon, specifically the Lunar Geophysical Network \citep{weber_scientific_2021}. One scientific goal would be to explore the lunar crust in more detail. Since the Apollo age, it is known that significant parts of it are KREEP terrains, understood to be mantle material from an overturn in an early lunar magma ocean. These can be distinguished spectrally from orbit, but should produce a significant imprint in crustal thickness. The crustal density models of the moon are actually those of highest resolution in the whole solar system, due to data from the GRAIL mission \citep{wieczorek_crust_2013}, but non-unique with respect to the average thickness of the crust, as well as the inner-crustal layering. Receiver function analysis to detect layering in the crust, was limited by the infamously strong scattering in lunar seismograms, but also by the low performance of the Apollo horizontal component seismometer. 

A future seismic network on the moon would consist of state of the art three-component seismometers, to better detect shear waves and converted phases. Central nodes could be equipped with very high fidelity sensors \citep{kawamura_autonomous_2022}, based on optical readout or superconducting gravimeters to directly observe normal modes. These would have resolution on the structure around the chemical layer above the core-mantle boundary and a potential inner core. Both features were seen in the analysis of \citet{weber_seismic_2011} and are of high importance to formation models not only of the Moon, but also of Earth, since the giant impact model derives the Moon from the Earth an a mars-sized impactor.
\subsection{Seismicity}
The lunar seismicity is famously divided into two families: the shallow and deep moonquakes. While the shallow moonquakes happen independently of another (their distribution in time is a Poisson process), the deep moonquakes follow a tidal cycle. The seismogram signals are so similar, that several clusters can be identified, in which moonquakes of similar focal mechanism repeat. This waveform similarity has been used to identify new moonquakes in the noise by template matching \citep{bulow_new_2005}.  In general, all moonquakes show a very high corner frequency compared to their moment magnitude \citep{oberst_unusually_1987}. The peaked response curve of the Apollo seismometers makes the estimation of the absolute event magnitude challenging \citep{kawamura_evaluation_2017}, but most estimates agree that the lunar seismicity is about four orders of magnitude below the terrestrial one \citep{knapmeyer_working_2006}.  
The seismicity of the Moon is relatively well known after nearly eight years of registrations by the Apollo stations. New mission concepts even rely on the recognition and re-use of source clusters that were identified in the Apollo data.
The Apollo data is however of limited value when concerning the far side of the Moon, where it is unknown if it is entirely aseismic, or if the core and partial melt layer on top of the core mantle boundary absorb all seismic waves from far side sources. Specifically the polar regions and the far side might harbour more regional tectonic activity that has been impossible to observe

\subsection{Mission perspectives}
The good accessibility of Moon, with launch windows almost every month means that mission complexity is significantly reduced. The NASA CLPS program tries to leverage this by buying delivery of scientific instruments to the lunar surface from commercial companies without prescribing a mission architecture. It is therefore likely that very different designs of seismic stations and network will be deployed over the next decade, starting with the Farside seismic suite (FSS), which is to land near Schr\"odinger crater in 2025 \citep{panning_farside_2021}. Among the main challenges are the high amount of radiation (essentially equivalent to interplanetary space, with the chance for solar extremes), the high temperature difference between day and night and the 14-day long cold night. The two latter require significant battery capacity to ensure heating and therefore survival of a lander through the night. The Apollo 11 EASEP station is an example for instrumentation that did not survive the lunar night.

A lunar geophysical network of multiple nodes, each equipped with a seismometer \citep{weber_scientific_2021} has been mentioned as a New Frontiers Class mission in the Decadal Survey of 2010 and 2022. The currently planned human landings during the Artemis program would allow to bring significant amounts of scientific payload with them, specifically, if the lander is a SpaceX "starship". However, at the time of writing, the specifics of the Artemis program, the exact landing concept and therefore, the scientific program are not yet defined.

\section{Mars}
\begin{figure}
    \centering
    \includegraphics[width=0.7\textwidth]{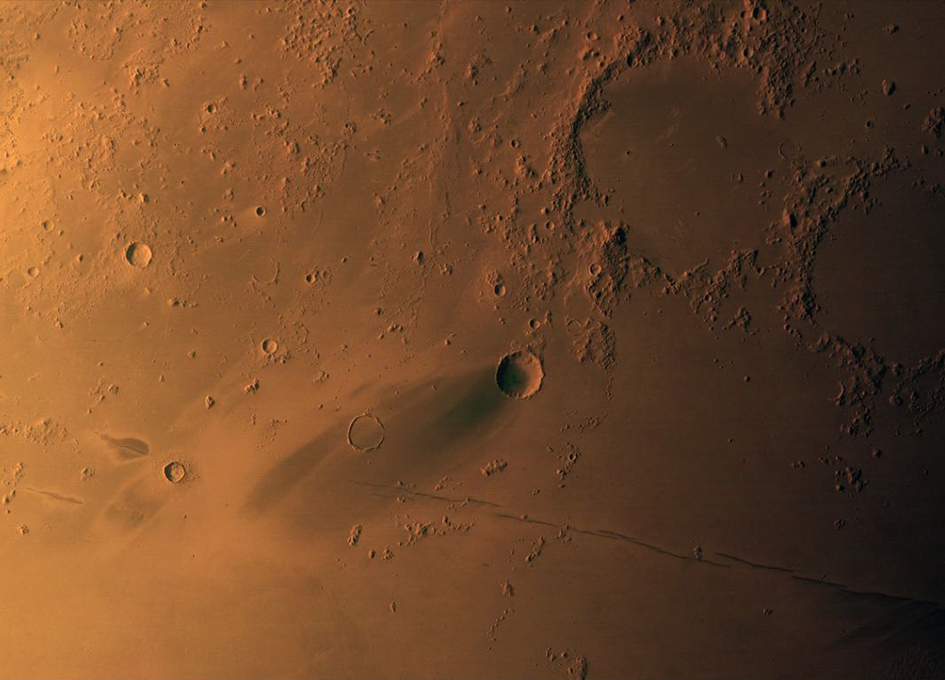}
    \caption{Image of South Western Elysium Planitia on Mars, taken on March 15, 2021, by the digital exploration camera (EXI) of the UAE Hope probe. The horizontal lines in the lower part of the image are the grabens of Cerberus Fossae, the source of most marsquakes recorded by the InSight mission so far \citep{giardini_seismicity_2020, perrin_geometry_2022, zenhausern_lowfrequency_2022}. }
    \label{fig:mars_hope}
\end{figure}
\begin{figure}
    \centering
    \includegraphics[width=\textwidth]{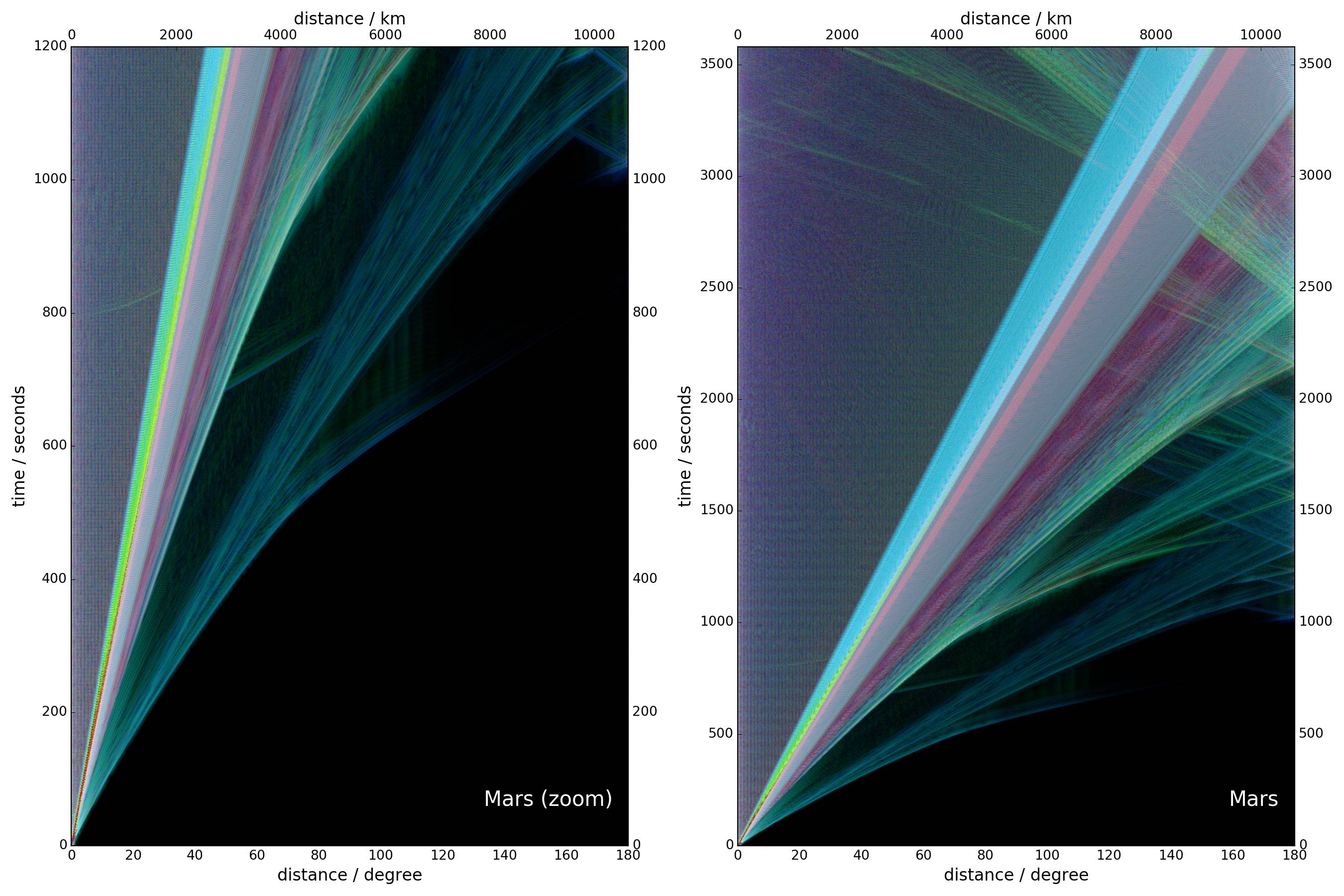}
    \caption{Global seismic wavefield stack for Mars, using the interior model InSight\_KKS21\_GP of \citet{stahler_seismic_2021-1}}
    \label{fig:gs_mars}
\end{figure}
\subsection{Potential scientific goals}
Apart from the Earth and the Moon, Mars is the only planet on which a seismometer has been operated for an extended period of time. The InSight mission \citep{banerdt_initial_2020} has shown that Mars is tectonically active, with a seismicity above that of the moon; similar to quiet intraplate regions on Earth.

Post-InSight seismic investigations could focus on one of three aspects:
\begin{itemize}
    \item InSight's ability to observe long period signals (below 50 mHz) was limited by the surface installation on loose sand. This prohibited observation of tidal deformation at the period of the Phobos orbit, which would have added a strong constraint on the rheology of the mantle \citep{van_hoolst_tidally_2003-1, lognonne_seis_2019}. A future long period seismic observatory would therefore have to be installed at least on bedrock to improve coupling to the ground at long periods. This installation would require either guided landing on exposed bedrock or a rover to deploy the seismometer on suitable ground in some distance of the lander.
    A second noise source was the tether, the rigid connector from SEIS to the lander, which contained 80 analog channels for scientific and housekeeping data \citep{zweifel_seismic_2021-1, hurst_resonances_2021, scholz_detection_2020}. A future mission would benefit from a much thinner tether, possibly only for power transmission, while data is transmitted wireless or on only two wires in serial digitized form.
    The long period background noise could be further reduced by burying the whole sensor assembly. This however would require a careful selection of a site where burial is possible to a depth of significantly compacted ground.
    \item While first layered models of the crust and mantle, including the radius of the core were obtained by InSight using a single station only \citep{duran_seismology_2022, khan_imaging_2021, knapmeyer-endrun_thickness_2021, stahler_seismic_2021-1}, many unknowns remain: The interior of the core has not been observed yet (e.g. by SKS waves), and even though tentative observations of Pdiff were made \citep{horleston_far_2022}, the lowermost mantle is constrained only very sparsely by seismic data. The detection of magnitude 4 quakes in Southern Tharsis \citep{horleston_far_2022} suggests that much more small seismicity could be present there, unobservable from InSight's location. At the same time, \citet{plesa_seismic_2021} showed that geodynamical models predict significant lateral variations in seismic wave speed, potentially higher than on Earth. 
    A global network of 4-6 seismometers, combined with other geophysical sensors could observe the full global seismicity, seismic phases over a wider distance range, and potentially also the above-mentioned three-dimensional structures in the Martian interior.
    \item InSight observed localized tectonic activity near the Cerberus Fossae graben, in contradiction to existing models of wide-spread compressive stress from lithospheric cooling \citep{phillips_expected_1991, knapmeyer_working_2006}. This observation has strong implications for the general mechanisms of tectonic activity on terrestrial planets. To better understand the mechanisms, it would be worth while to locate marsquakes precisely in one of the active systems, for which a multi-station seismic network is necessary. Due to the low intrinsic seismic attenuation, energy above frequencies of 1 Hz is transmitted well and observable, which allows to use light-weight, short period instruments, such as the InSight SP-sensor \citep{stahler_cerberus_2022}. 
    
\end{itemize}
\subsection{Mission perspectives}
Landing on Mars has been executed successfully nine times by NASA and once each by the Soviet Union (Mars 3 on December 2, 1971, \citet{perminov_difficult_1999}) and the Chinese Space Agency (Zhurong on May 14, 2021). All these landers were fundamentally operating a combination of a heat-shield and parachute for initial deceleration. The Mars Exploration Rovers Spirit and Opportunity used airbags for final approach and landing, while the large Perseverance and Curiosity rovers were lowered to the ground from a separate spacecraft, colloquially termed sky crane. The Discovery class missions InSight and Phoenix used hydrazine retrothrusters for final deceleration. These techniques are currently considered to place a lower limit of about 200 million USD on any mission, before scientific instruments are even considered, requiring at least a Discovery class budget for future landed missions. 

A potentially cheaper option is to use penetrators, i.e. spacecraft that are landing with terminal velocity and decelerate either by penetrating up to a few meters into ground or by having a deformable front part. As described in \citet{lorenz_planetary_2011}, the concept has been proposed several times over the last decades, but was never successfully executed. However, it must be noted that no penetrator mission failed during landing; Mars-96 was lost due to failed upper stage separation after launch, while DS-2 likely got lost with the mother spacecraft Mars Polar Lander. Overall, Mars is the location suited best for simple penetrator missions, since the atmosphere can fulfill initial deceleration and - more importantly for a low-cost mission - attitude control, removing the need for active propulsion.
A semi-hard lander concept has recently been proposed by engineers from JPL under the name SHIELD \citep{barba_access_2021}. It would utilize a deployable drag skirt of 2 meter diameter for deceleration to obtain a low ballistic coefficient and therefore terminal velocity of $<70$~m/s without the use of a parachute. At least four of such landers could be stored in a single Falcon-9 sized payload fairing and would therefore allow to build a seismic network on Mars.
During landing, instruments, including the seismometer would have to survive deceleration of up to 2000~g, depending on surface character. The JAXA seismometer planned for the Lunar-A mission \citep{mizutani_lunar_1995, shiraishi_present_2008} is specified for 5000~g and would therefore be a candidate instrument. The seismometers of the Ranger 3/4/5 missions was tested up to 3000g \citep{lehner_seismograph_1962}. For a free-fall landing on the Moon, it was encapsulated in a balsa wood impact limiter sphere and submerged in freon. Other instruments, such as the InSight SP seismometer could be hardened in a similar way during flight and landing.

A ultra-high sensitivity mission would require a soft lander, likely in combination with a rover to reach a suitable installation site after landing. While such a mission is well within the technical possibilities of NASA and likely also CNSA and ESA, it would likely require a flagship class budget for landing and operations and could therefore only be executed as part of a larger rover mission. However, NASA and ESA are currently executing the Mars Sample Return campaign, involving the Perseverance rover and likely 3 more launches until 2030. Given the size of the mission and its considerable strain on the budget of both involved agencies, it is unlikely that any additional scientific payload will be added to it, to avoid mission and budget creep. The delay of the ESA ExoMars landing due to the stop in collaboration with Roscosmos will further affect all Western Mars missions.

As a final word, it should be noted that the landscape of possibilities on Mars is likely to change, if various private space companies, foremost SpaceX, succeed in constructing reusable high performance launch vehicles, that would increase mission cadence and lessen weight limitations. However, launch cost is not at all a constraining factor typically and other factors (e.g. the availability of a high bandwidth communication network) limit mission design on Mars.

\section{Phobos and Deimos}
\subsection{Potential scientific goals}
For the two Martian moons, it is not even known whether they consist of consolidated material or form rubble piles \citep{le_maistre_signature_2019, dmitrovskii_constraints_2022}. If they form rubble piles, seismic wave propagation in a usual sense is not possible and the strongest signal will instead be the normal modes of the moons. A more consolidated object however could be propagating seismic waves efficiently; due to the small size with high amplitudes even for small quakes. The high thermal gradient and strong tidal signal from Mars is likely to trigger a significant amount of small phobos- or deimosquakes, if the moons as a whole are consolidated enough.
The rigidity of the uppermost surface layer could be estimated from the deceleration of a lander, information on mechanical properties of the soil surface will also be gained from the JAXA MMX rover, which will land and drive on the surface of Phobos.

\subsection{Mission perspectives}
The launch windows for Phobos and Deimos are identical to those of Mars, yet landing requires an entirely different skill set. A sample return mission to Phobos, MMX (Martian Moon eXplorer) is planned by JAXA for launch in 2024 \citep{kuramoto_martian_2022}, building on their extensive expertise in missions to asteroids. The mission will not carry a seismometer, but during the landed time windows, the spacecraft's IMU will listen for ground vibrations. Given that the lander stays on the surface for only a few hours and performs sampling operations during that time, the detection of a phobosquake would be a lucky event.
The mission will perform a geodetic experiment, mapping the shape of Phobos using imaging and LIDAR and combining it with Doppler radio tracking of the spacecraft to obtain gravity coefficients from which the interior can be inferred \citep{matsumoto_mmx_2021}.

\section{Ceres}
\begin{figure}
    \centering
    \includegraphics[width=0.5\textwidth]{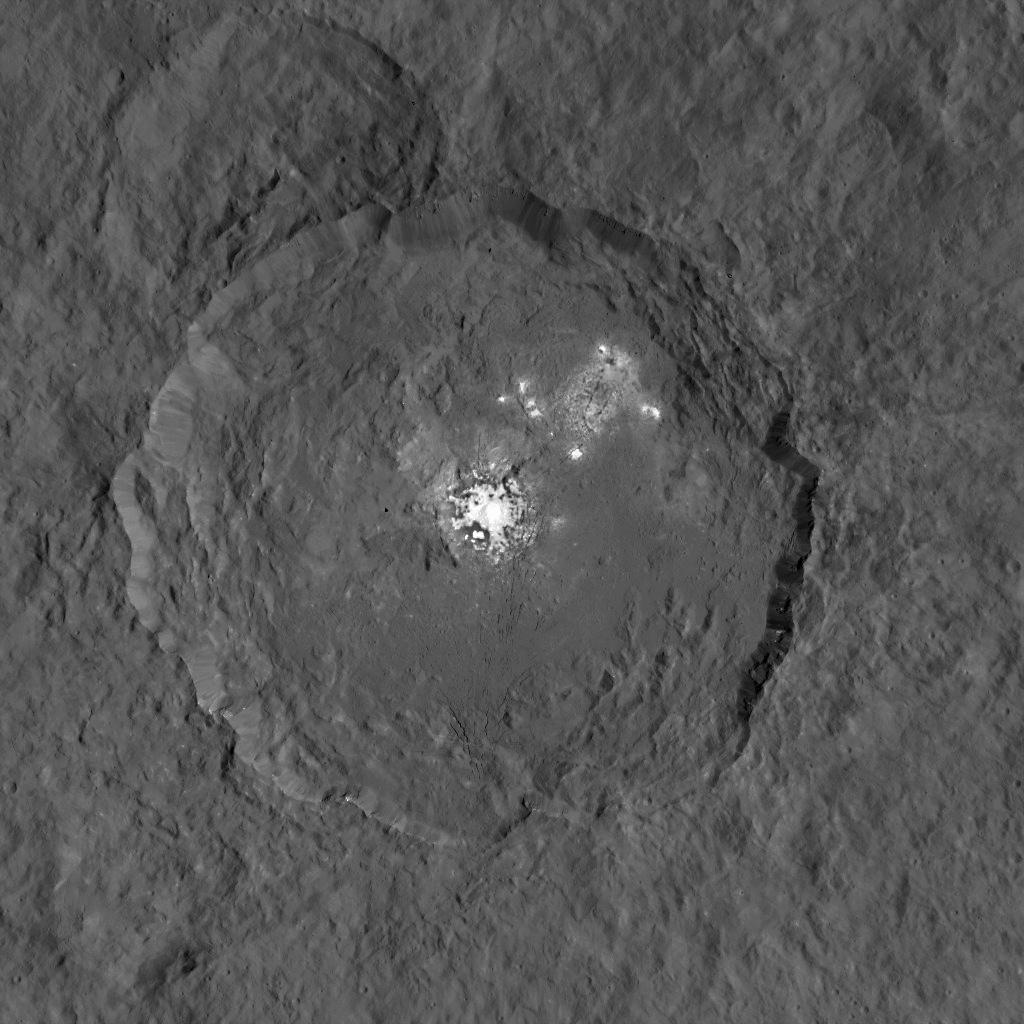}
    \caption{Occator crater on Ceres, the proposed landing site for a NF-class Ceres sample return mission \citep{castillo-rogez_concepts_2022}. The bright spots are understood to be young carbonate salt deposits from subsurface brines. A seismometer could help to constrain the current tectonic activity and from it the deposition rate. The view is a composite of two DAWN images, avoiding over-exposure of the bright spots. Image credit: NASA/JPL-Caltech/UCLA/MPS/DLR/IDA, PIA19889}
    \label{fig:ceres}
\end{figure}
\subsection{Potential scientific goals}
Ceres shows trace of a subsurface ocean, which would contain a significant amount of the water of the asteroid belt. Surface morphology indicates that this ocean is indeed partially liquid and forms yet another ocean world. A seismic mission could explore the depth of the ice/ocean interface, the thickness of the ocean, but also the homogeneity of the ice layer above. A warm ocean would result in slushy layers at the bottom of the ice that would significantly increase attenuation.

Since the planet has no moons, the tidal dissipation is insignificant. Yet, the NASA Dawn mission observed signs of recent cryovolcanism \citep{ruesch_cryovolcanism_2016}: The Ahuna Mons topographic feature is 4 km high at a width of 16 km and only lightly cratered. The age is estimated to be $210 \pm 30$ million years, i.e. very young in the context of a small planet. It is debated whether the dominant tectonic process on Ceres is solid state convection inside the icy crust, leading to dome formation \citep{bland_dome_2019} or instead global contraction, as evidenced by ubiquitous thrust faults \citep{ruiz_evidence_2019}. It is well possible that these thrust faults were indeed created in an early stage of the planet's formation and are thus fossils of the planet's ancient tectonics. The latter seems to be the case on Mars, too, where no seismicity could be attributed to thrust faults, lobate scarps or wrinkle ridges so far.
Following the example of Mars, where Cerberus Fossae, one of the youngest surface features is also the most seismically active, Occator crater would be a prime target. The crater shows bright spots (faculae), interpreted as deposits of salts from eruption of brines \citep{nathues_recent_2020, schenk_impact_2020} (see fig. \ref{fig:ceres}). If these brines are deposited by an endogenic process, it would also lead to contemporary seismic activity, that could be picked up even by a short period seismometer.

A broadband Ceres seismometer would investigate the global distribution of tectonic activity. Since high resolution orbital images exist from the Dawn mission, at a resolution of 35 m, attribution to individual fault systems or centers of cryovolcanism would be possible. Assuming large enough Ceresquakes, the seismic data could be used to investigate layering of the planet, including the depth of the ocean.
\subsection{Seismicity}
The seismicity of Ceres is unknown. Given the absence of a partner object for tidal forcing and low interior heat flow, it is likely to be low. The presence of recently deposited brines \citep{nathues_recent_2020} suggests that seismicity driven by interior processes might be present. Due to the location of Ceres in the Asteroid belt, meteorite impacts are a likely seismic source.
\subsection{Mission perspectives}
Ceres can be reached relatively easily, typically with a Mars flyby, leading to a 3-4 year trajectory for orbiter missions. The Dawn mission used solar electric propulsion, leading to a 6 year trajectory, including a 400 day research trip to asteroid Vesta. 

In the aftermath of the Dawn mission, lander missions have been proposed; due to the low gravity, even sample return missions are feasible. An interesting concept is proposed by \citet{castillo-rogez_concepts_2022}, in form of a electrically propulsed orbiter that is able to land and relaunch as a whole, returning samples to Earth. The mission is supposed to collect samples of carbonate salts as well as the darker reference surface materials and return it to Earth at $\leq 20$\textdegree C to prevent alteration. The mission would benefit from investigating recent tectonic activity in Occator crater, to determine the samples' context. Unfortunately no seismometer is currently foreseen. A slightly modified version of this mission was proposed as a candidate for a New Frontier class mission in the Decadal Survey 2023-2033 \citep{decadal_2023}, still without seismic measurements.

A tip to the hat goes to the student participants of the Alpach summer school, who proposed a more classical dual orbiter/lander sample return mission, likely in the cost range of a NASA flagship mission or an ESA L-class mission \citep{gassot_calathus_2021-1}.

\section{Jupiter and Saturn - the giant planets}
\begin{figure}
    \centering
    \includegraphics[width=0.6\textwidth]{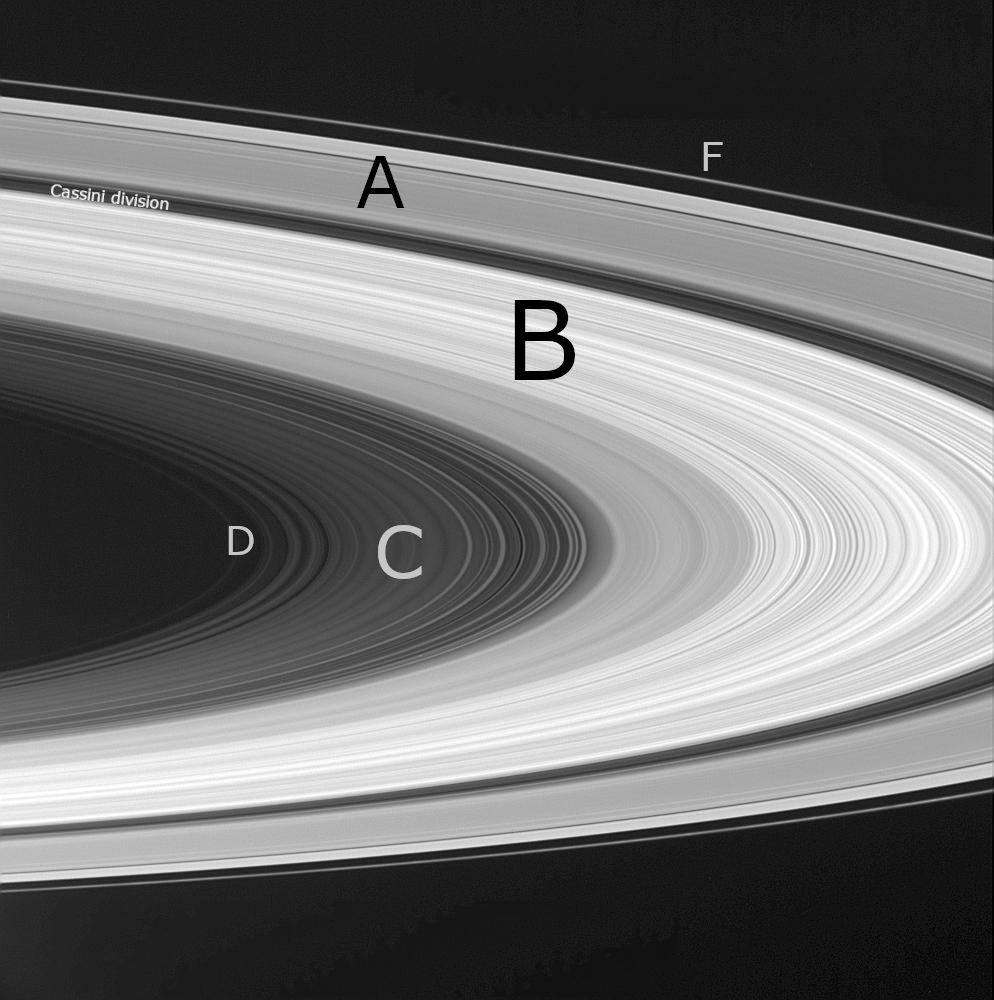}
    \caption{Cassini image of Saturn's rings taken by the ISS infrared camera. Major subdivisions are labeled by the author. Kronoseismology is done using density waves in the broad, but faint C-ring. Image credit: NASA/JPL/Space Science Institute, PIA06536}
    \label{fig:saturn_ring}
\end{figure}
\subsection{Potential scientific goals}

For this topic, see the excellent overview in \citet{gaulme_giant_2015}. Seismology in the common sense, using a landed mass-and-spring sensor is of course impossible on a gas giant, instead measurements focus on long period deformation, detected by astronomical means. As all bodies, the giant planets have normal modes whose shape and frequency depends on their interior's elastic parameters. Since the bulk of the interior is in fluid state without shear modulus and yet gravity is significant given their size, the nomenclature of helioseismology is typically used. Compared to Earth, where the elastic moduli dominate as restoring force and gravity and self-rotation (the Coriolis force) can be treated as second order parameters, in the sun or the gas giants, the modes need to be treated separately by their primary restoring force \citep{guillot_giant_2015}.
\begin{itemize}
\item pressure modes (p-modes) are the closest analog to normal modes on Earth. The restoring force is the pressure field, or the gradient of the scalar potential $p=\nabla \phi$, analog to the case of the acoustic wave equation. Their sensitivity is typically constrained to the outer layers.
\item gravity modes (g-modes) have the gravity, or more precisely buoyancy as restoring force. They can therefore only form in convectively stable regions, where no density inversions exist and lateral density contrasts are low.
\item surface gravity modes (f-modes) are analogous to deep water gravity waves in the ocean, with the weight of surface vertical displacement as restoring force. Their sensitivity is therefore highly constrained to the outermost shell of the planet.
\item inertial modes (i-modes) can occur in rapidly rotating planets, such as Jupiter and Saturn, restored by the Coriolis force.
\end{itemize}
All in all, these modes create a complex overlapping picture of spectral peaks in the surface displacement and of course cross terms and coupling exist. Over time, four approaches have been considered to observe the normal modes of Jupiter remotely
\begin{itemize}
\item Variations in infrared brightness, caused by temperature perturbations from p-modes. A 1~m/s velocity field corresponds to 10~mK in temperature perturbation, visible in mid-infrared, which are difficult to observe, given the limited sensitivity of photometric sensors in this optical wavelength window. Because of these difficulties, no dedicated instrument has been developed so far. 
\item Spectroscopy of reflected light. This method has been improved considerably on instrument side since in response to the exoplanet detection campaigns. In 2011, the seismology-dedicated SYMPA Fourier spectro-imager detected radial modes of maximum amplitude of $49^{+8}_{-10}$~cm, at a frequency of $1213 \pm 50~\mu$Hz, with a mean large frequency spacing between radial harmonics of $155.3 \pm 2.2~\mu$Hz \citep{gaulme_detection_2011}, placing a weak constraint on the planets interior structure. Spectroscopy needs to take into account the large rotation contribution at the fringes of around $25$~km/s.
\item Photometry of reflected light. Here, brightness variations in reflected light are used. Compared to the other two methods, the signal-to-noise ratio is low, but sensors in visible wavelengths are widespread. However, the complex surface pattern of the Jovian clouds means that only specific modes can be observed. Another problem arose when trying to apply this method to Neptune with the extended NASA Kepler mission "K2", which observed the planet for 50 days at a 1-min cadence. No oscillations of Neptune could be detected (Rowe et al 2017). But it became apparent that oscillations of the Sun in the reflected light perturbed the signal (Gaulme et al. 2016).
\item Kronoseismology: The rings of Saturn are shaped by resonance with the interior modes of the planet. Density waves exist in all rings, but specifically in the innermost C-ring (see fig. \ref{fig:saturn_ring}), they  are excited by certain normal modes of the planet \citep{hedman_kronoseismology_2013}. This is observationally accessible via photometry of the exact ring pattern during stellar occultations during the Cassini mission \citep{fuller_saturn_2014} and led to the discovery of a diffuse, but stable stratified core of the planet \citep{mankovich_diffuse_2021}.
\end{itemize}
\subsection{Mission perspectives}
Using ring seismology, the interior models of Saturn cannot be refined any further, given the high resolution on density waves in the C-ring obtained by Cassini from radio occultation experiments. Future research would therefore focus on Jupiter. Gravimetric observations by Juno \citep{durante_peek_2021} detected gravity perturbations that are compatible with the presence of p-modes, which proved the existence and excitation of these modes, although they could not be further identified.

A Doppler imaging camera for Jupiter seismology was proposed as a payload for the ESA JUICE mission \citep{soulat_echoes_2011}, but ultimately not selected. Given that Juno, a general purpose mission is currently active in the Jupiter system and two flagship class orbiter missions (JUICE and Europa Clipper) are due to launch until 2025 (although both focused on the icy moons), the need for another large Jupiter orbiter has not been foreseen by the Decadal Survey. Instead, Doppler spectroscopy from Earth will likely be the only possible method for the foreseeable future. The success of SYMPA was followed on by the JOVIAL sensor \citep{goncalves_first_2019} and dedicated instruments to measure Jupiter's zonal wind speeds, e.g. PMODE-I on the AEOS 3.6 m telescope atop Mount Haleakal\=a, Maui, Hawai'i \citep{shaw_pmode_2022} are built. Observations of the normal modes with increased precision compared to SYMPA would however be fortuitous, possibly due to a large impact. To measure Jupiter's interior modes with high enough precision, multi-week continuous observation campaigns (at stable weather) would be necessary, which are likely only possible from Antarctica \citep{shaw_pmode_2022}. In the far future, it could also be done from a long-lived optical Jupiter observatory at Lagrange-point L1, as proposed by the Chinese space agency CNSA \citep{hsu_jupiter_2021}.

\section{Io}
\subsection{Potential scientific goals}
While Europa, Ganymede and Callisto are all icy moons, Io is a terrestrial planet without any surface ices \citep[or water absorption features whatsoever][]{smith_jupiter_1979}. The ice fraction on these moons actually increases with distance from Jupiter, which is one of the problems moon formation models face \citep{shibaike_galilean_2019}. In how far the interior structure and composition of Io is related to the structures of the other icy moon's rocky cores, and in how far these permit certain formation scenarios, can be understood only if the interior structures of them all are better resolved.
While Io's active volcanism (discovered by Voyager I, \citep{smith_jupiter_1979} is a fascinating mission goal in itself, the moon is also a perfect natural laboratory to understand tidal heating in multi-body systems. To quote the KISS report on Tidal heating: "The Io–Europa–Ganymede system is a complex and delicately built tidal engine that powers Io’s extreme volcanism and warms water oceans in Europa. Io’s gravity generates a tidal bulge within Jupiter, whose dissipation transfers some of Jupiter’s rotational energy into Io’s orbit, moving it outwards and deeper into a 2:1 eccentricity resonance with Europa. This in turn increases Io’s eccentricity, resulting in enhanced tidal heating. Ultimately, Jupiter’s rotational energy is converted into a combination of gravitational potential energy (orbits of the satellites) and heat via dissipation in both Jupiter and the satellites" \citep{de_kleer_tidal_2019}. The tidal heating is ultimately the cause of Europa's liquid ocean, which may be a permanent feature or periodic \citep{hussmann_thermal-orbital_2004}. Since tidal heating is the process creating the many ocean worlds in the Solar system \citep{nimmo_ocean_2016} and in other exoplanet systems, understanding it has significant consequences for habitability, as well as evolution of planetary systems.
Prior to the observations of Voyager I, \citet{peale_melting_1979} predicted that the deep interior of Io should be largely molten because of the tidal heating. \citet{schubert_internal_1981}, in explicit contradiction, proposed a thin, molten layer between the crust and a solid interior and an iron core. The thickness of the molten layer is estimated to be at least 50 km, with a melt fraction exceeding 20\% \citep{khurana_evidence_2011}. \citet{van_hoolst_librations_2020} even consider a magma ocean possible. The size of the core is somewhere between 19\% and 50\% of Io's radius \citep{anderson_primary_2001}. Seismological experiments could determine layer thicknesses and core radius, and, via shear modulus and attenutation, constrain the melt fraction in the asthenosphere.

\subsection{Seismicity}
As with all other large moons of the giant planets, Io's seismicity is not known. However, the planet shows obvious surface tectonic activity, which implies large strain and regular brittle failure, i.e. ioquakes. A popular approach to estimate seismicity has been to use the (relatively well known) tidal dissipation and assume that a certain part of that energy is released seismically. \citet{hurford_seismicity_2020} estimated the ratio between tidal dissipation and seismic energy over an orbital cycle for the Earth's moon to be 0.0017 and assumed that this ratio would be a good first order estimate for other tidally active worlds. From this assumption, they found that Io would release an annual seismic moment on the order of $5\cdot10^{19}$ Nm/a (see table \ref{tab:tidal}, where the values are per ten orbital cycles). Assuming that the largest possible ioquake releases 70\% of the available moment (in analogy to Shallow Moonquakes), this implies that one to two magnitude 6 ioquakes would occur per month - a rate comparable to that of the Earth.  The assumptions behind this scaling analysis may of course be simplistic, but it's nevertheless highly plausible that Io may be the most seismically active solid body after Earth. The existence of ridges structures oriented according to the directions of tidal stresses supports that tidally driven tectonism could be active \citep{bart_ridges_2004}.

While the existence of tectonic events is currently hypothetical, the volcanic activity of Io is obvious and documented photographically \citep{smith_jupiter_1979}. One can thus expect to record seismic signals of the kind known from volcanos on Earth, i.e. distinct transient events as well as a more or less continuous tremor. Optical observation of volcanic centers on the surface could provide epicenter information and thus support the construction of travel time curves and the inversion for interior structure.

\subsection{Mission perspectives}
Despite the spectacular surface colors, landing on Io is actually not particularly dangerous, since the surface volcanism is locally constrained. A larger problem is Io's place very deep in Jupiter's gravity well, which would require considerable \deltaV~reserves and a launcher of SLS or Falcon Heavy class. The location in Jupiter's radiation belt poses a major design driver for lander electronics - contrary to orbiter missions which can minimize the radiation dose by performing only brief, repeated flybys, as in the IVO mission profile \citep{adams_io_2012, mcewen_io_2014}, while a lander would need dedicated shielding of electronics. 

A more realistic option for seismology may therefore be an orbital detection of seismic deformation using Interferometric Synthetic Aperture Radar (InSAR) during multiple flybys or deployment of a few retroflectors that can be queried with a laser on an orbiter. A more realistic option for seismology may therefore be an orbital detection of coseismic deformation using Interferometric Synthetic Aperture Radar (InSAR) during multiple flybys, a technique which is routinely applied to significant Earthquakes, or deployment of a few retroflectors that can be queried with a laser on an orbiter (Laser vibrometry, see \citet{de_kleer_tidal_2019}, and references therein). Via the latter, normal modes could be observed to constrain the deep interior of the planet, while the former would mostly serve to understand shallower lithospheric strength.

\begin{table*}[!ht]
    \centering
    \resizebox{\textwidth}{!}{
    \begin{tabular}{|l|l|l|l|l|l|l|l|l|l|l|l|l|}
    \hline
         & $m_{\textrm{parent}}$ & Period & $R$ & $a$ & $e$ & $k_2$/$Q$ & $E_{\textrm{T}}$ & $\sum M_0$ & $M_{\textrm{w}}$ & $M_C$  & $M_{\textrm{w}}$  \\ \hline
         & [kg] & [days] & [km] & [km] & [\%] &  & [J] & [Nm]  &  &   [Nm] &  \\ \hline
         Io & $1.90\cdot10^{27}$ & 1.769 & 1821.6 & $421,700$ & 0.41 & 0.015& $1.43\cdot10^{20}$ & $2.3\cdot10^{18}$ & 6.2 & $1.7\cdot10^{18}$ & 6.1 \\ \hline
         Europa & $1.90\cdot10^{27}$ & 3.551 & 1560.8 & $670,900$ & 1 & 0.0054 & $8.7\cdot10^{18}$ & $1.5\cdot10^{17}$ & 5.4 & $1.0\cdot10^{17}$ & 5.3 \\ \hline
         Titan & $5.68\cdot10^{26}$ & 15.945 & 2575 & $1,200,000$ & 2.88 & 0.004 & $1.6\cdot10^{18}$ & $2.7\cdot10^{16}$ & 4.9 & $1.9\cdot10^{16}$ & 4.8 \\ \hline
         Moon & $5.97\cdot10^{24}$ & 27.3 & 1737.2 & $384,399$ & 5.5 & 0.0012 & $5.0\cdot10^{16}$ & $8.0\cdot10^{14}$ & 3.9 & $4.9\cdot10^{14}$ & 3.7 \\ \hline
         Enceladus & $5.68\cdot10^{26}$ & 1.37 & 252 & $237,948$ & 0.47 & 0.0036 & $6.3\cdot10^{15}$ & $1.0\cdot10^{14}$ & 3.3 & $7.5\cdot10^{13}$ & 3.2 \\ \hline
        Earth/Lunar & – & 1 & – &  & – & – & $7.2\cdot10^{16}$ & $1.2\cdot10^{15}$ & 4 & $8.5\cdot10^{14}$ & 3.9 \\ \hline
         Mars/Solar & – & 1.03 & – &  & – & – & $8.9\cdot10^{14}$ & $1.5\cdot10^{13}$ & 2.7 & $1.\cdot10^{13}$ & 2.6 \\ \hline
         Mars/Phobos & – & 0.32 & – &  & – & – & $9.2\cdot10^{11}$ & $1.6\cdot10^{10}$ & 0.7 & $1.1\cdot10^{10}$ & 0.6 \\ \hline
         Mercury/Solar & – & 58.65 & – &  & – & – & $1.2\cdot10^{15}$ & $1.2\cdot10^{15}$ & 4 & $8.4\cdot10^{14}$ & 3.9 \\ \hline
    \end{tabular}
    }
    \caption{Estimated tidally-induced seismic moment released over ten orbital cycles. The table is slightly modified from \citep{hurford_seismicity_2020} and references therein. Note that this table assumes that the ratio of tidal dissipation to seismically released energy found on the moon (0.0017) is applicable to the other worlds. Also note that ten orbital cycles are 18 terrestrial days on Io, but 6700 days on Mars.}
    \label{tab:tidal}
\end{table*}

\section{Europa}
\begin{figure}
    \centering
    \includegraphics[width=\textwidth]{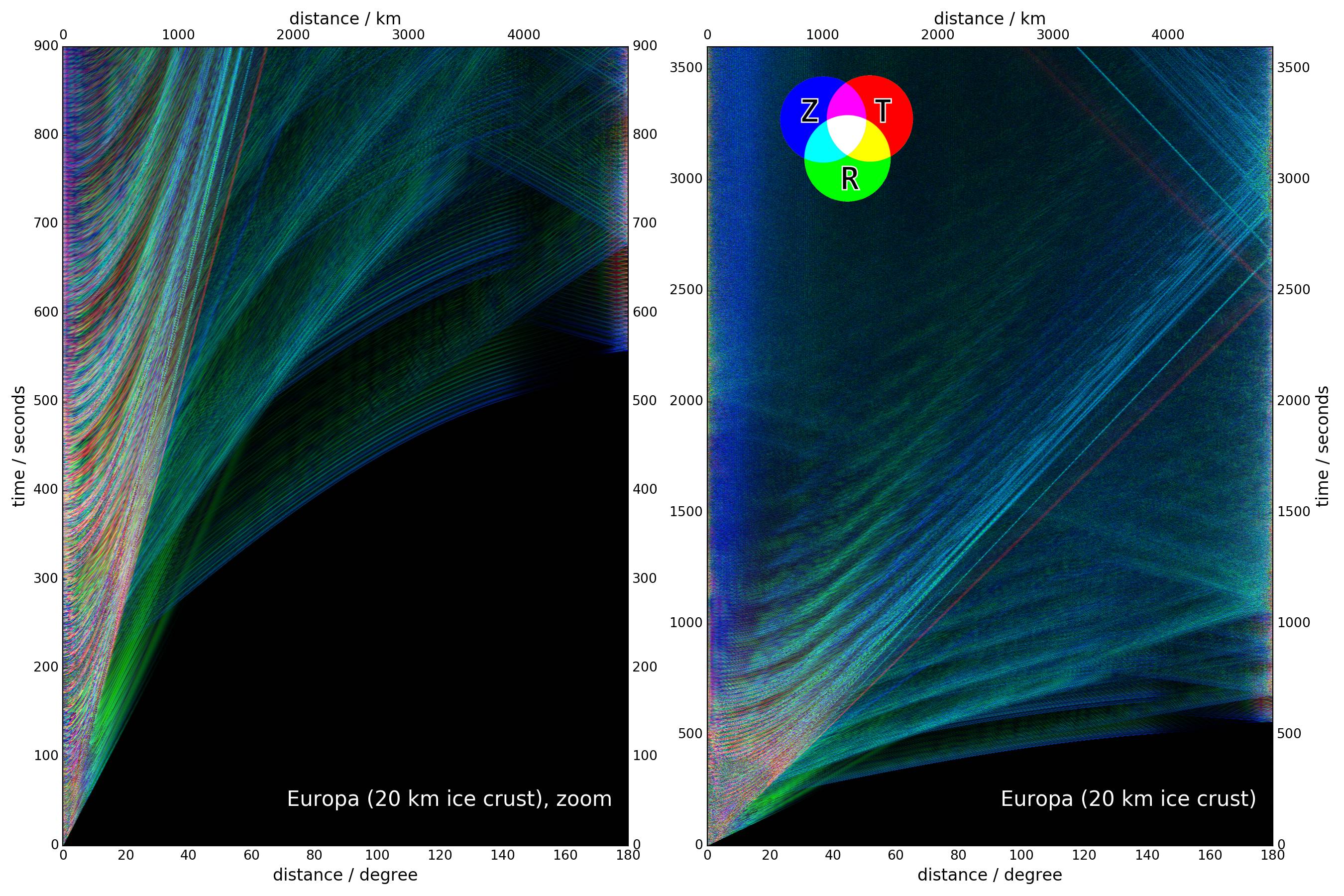}
    \caption{Global seismic wavefield stack for Europa, using an interior model of \citet{vance_geophysical_2018} with a crustal thickness of 20 km.}
    \label{fig:gs_europa}
\end{figure}
\subsection{Potential scientific goals}
A liquid subsurface ocean on Europa was first predicted from the internal energy budget by radioactive decay and tides \citep{lewis_satellites_1971, cassen_is_1979}, and later supported by flyby measurements during the Galileo mission \citep{schubert_interior_2004}. The thickness of the ice layer above the ocean is related to composition and temperature of the water; measurements of said value would therefore constrains these parameters, which have strong implications for habitability of the ocean. The ESA JUICE mission, as well as the NASA Europa Clipper will therefore both carry radars to measure the thickness during multiple flybys \citep{bruzzone_rime_2013, phillips_europa_2014}. At this stage, the attenuation of electromagnetic waves by ice under Europa conditions is not well known, and the detection of a reflection of the ice bottom by either of the mission's radars is likely, but not certain \citep{eluszkiewicz_dim_2004, aglyamov_bright_2017}.
A surface-deployed seismometer would be sensitive to ice thickness via 3 routes:
\begin{itemize}
\item An layer above a fluid half space forms a specific seismic phase, termed Crary phase after it's first descriptions on floating arctic ice \citep{press_propagation_1951, crary_seismic_1954}. The phase is an almost monochromatic, radially horizontally polarized superposition of SV reverberations. Its central frequency is $
f_{\mathrm{Cr}} = \frac{v_{\mathrm{S}}}{2d \sqrt{1 - \left(\frac{v_{\mathrm{S}}}{v_{\mathrm{P}}}\right)^2}},$, where $d$ is the ice thickness and $v_{\mathrm{S}}, v_{\mathrm{P}}$ are S- and P-wave speeds respectively. For the range of ice thicknesses predicted for Europa (5-30 km) \citep{vance_vital_2018}, $f_{\mathrm{Cr}} $ would be between 0.11~Hz and 0.44~Hz, i.e. well-observable by a short period seismometer. This has made it a prominent candidate for ice thickness determination since 20 years \citep{kovach_seismic_2001}; the phase is also well-determinable in synthetic seismograms \citep{stahler_seismic_2018}, but its robustness against heterogeneous ice of varying thickness is at this stage not known.
\item Since the ocean cannot propagate shear waves, S-waves and specifically horizontally polarized SH waves will be reflected almost fully at the ice-ocean interface. Any seismic signal should therefore contain strong reverberations of shear waves, whose traveltime $T_{\mathrm{rev}}$ would be directly proportional to ice thickness $d$: $T_{\mathrm{rev}}=2d/v_{\mathrm{S}}$. The observability of direct reverberating phases would be affected by seismic attenuation, which reduces shear wave amplitude strongest. However, the reverberating waves would also be present in the ambient seismic noise of Europa and could be retrieved by autocorrelation \citep{stahler_seismic_2018}.
\item The ice thickness places an upper limit on the period of Rayleigh waves. If the ice is on the thinner side of previous estimates (below 10 km), this limit would be below 5 seconds, i.e. potentially observable by a short period sensor \citep{panning_long-period_2006, stahler_seismic_2018}. 
\end{itemize}
All in all, the different potential observables provide a certain redundancy in determining the ice thickness using seismic methods, which is one of the reasons, why a seismometer has been an element of the Europa lander mission concept, that is currently awaiting NASA adoption and funding \citep{hand_report_2017}. 

This mission described several other science goals of a seismometer, specifically to observe water or brine lenses within the ice. Seismic methods are usually good at detecting heterogeneities at depth, but as it was ruthlessly pointed out by \citep{grimm_magnetotelluric_2021}, the elastic impedance contrast between mushy ice and liquids is not particularly strong and determining even a "1D" layered seismic velocity profile below a lander is by no means a trivial process. Yet, the non-uniqueness of geophysical observations in a single station is even worse for electromagnetic, specifically potential-based methods. Also, \citet{hobiger_shallow_2021} demonstrated using InSight data that long-term observation of ambient vibrations at a single location can constrain even complicated subsurface structure, using geological context information, if available.

The habitability of Europa's ocean would be increased by transport of surface material into the ocean, since the Jovian radiation oxidizes surface material, creating a potential energy source for primitive life, if transferred back into the ocean. Subduction in the ice has been proposed based on the geomorphology of linea on Europa's surface \citep{prockter_folds_2000}, but has not been observed in-situ. As InSight demonstrated on Mars, locations of seismic sources can be determined well using a single seismometer and shed light into the dominant tectonic process on a planet. Subduction zones on Earth cause well-observable continuous seismicity even in between the largest events, so a seismometer deployed close to a linea should be able to pick up its signature. This would confirm imaging data from orbit in one location and would allow inference over the whole planet (and in similar locations on Enceladus).

A separate question of high importance for habitability is the geological activity of the sea floor. The mid ocean ridges on Earth provide habitable environments independent of sunlight and sub-ocean volcanism on Europa could play a similar role. Such activity, even in the past could be inferred from the chemistry of the dark patterns on Europa's surface, but europaquakes in the silicate crust or mantle would be a unique indication of current activity. \citet{marusiak_seismic_2022} estimated that rocky europaquakes above magnitude 5 would be observable by a surface seismometer of the kind proposed for the Europa lander. Such magnitudes would be quite significant and are at the upper end of what is observed on Earth's mid-ocean ridges directly (not counting the adjacent transform faults, which often produce earthquakes of 7 and higher), but by no means impossible.

Finally, the sound speed in the ocean itself is affected by the ocean's chemistry. A final seismic velocity model of the planet at the end of the mission would be contigent of the ocean's salt content. \citet{duran_seismology_2022} demonstrated a fully consistent inversion of different seismic observables with other geophysical, as well as mineralogical data in a thermodynamically consistent model for Mars. Such an effort could be the final result of an observation campaign on Europa as well, delivering uncertainty limits on composition of the ocean, the thickness of the various layers and the wave speeds in the silicate interior.

\subsection{Seismicity}
\citet{panning_expected_2018} estimated the seismicity rate of Europa based on the available energy from tidal dissipation to be between $10^{16} - 10^{18}$~Nm/a, which is above the value of $10^{15}-10^{16}$~Nm/a observed for Mars \citep{banerdt_initial_2020}, but 5 orders of magnitude below the Earth. Assuming this value is correct, a few dozen europaquakes from the icy crust should be observable over a month by a Europa seismometer as defined in \citep{hand_report_2017}. Assuming that the ratio of tidal dissipation to seismic moment is about the same on all tidally active moons in the solar system \citep{hurford_seismicity_2020}, Europa would be the second most active moon after Io (see table \ref{tab:tidal}).

\subsection{Mission perspectives}
Europa is challenging to land and operate on. Landing needs to be entirely propulsive, due to the lack of an atmosphere and Europa is deep in the gravity well of Jupiter. Since Europa is in the distance of highest radiation within Jupiter's magnetic field, a surface mission requires extensive and heavy metal shielding, which \textit{worsens} the effect of the high-$\delta V$ landing. Within realistic weight constraints, a surface mission is limited to a duration of a few weeks and a scientific payload of tens of kg. The Europa lander concept \citep{hand_report_2017} managed to fit a seismometer into these constraints. The instrument would be similar in performance to the InSight SP seismometer \citep{pike_silicon_2016, lognonne_seis_2019}, with the goal of listening to body waves of europaquakes in the ice. Observation of long period surface waves, flexural ice modes or normal modes of the whole planet would likely not be possible with an instrument of this sensitivity. The observability of seismic waves from quakes in the silicate mantle would highly depend on their magnitude, but would be made less likely by the presence of soft layers on the seafloor or the ice bottom \citep{marusiak_seismic_2021}. In general, operation of a warm lander on ice will pose challenges to coupling of seismic sensors due to melting and tilting \citep{marusiak_detection_2022}.

A spacecraft in orbit at Jupiter distance can still operate on solar panels, but their size would be prohibitive for a landed mission. Power would therefore have to come from an RTG, or for a short-lived mission from high-power-density batteries.

The Europa Lander concept is the last iteration of a 3 decade long process of missions to explore the planets surface and interior. The ups and downs of this history are excellently described in \citet{brown_mission_2021}. At this stage, it is a well-developed concept with a surface lifetime of 60-90 days. Yet, it would be the most expensive planetary robotic mission ever executed by NASA, which is why the Decadal Survey 2023-2033 did not recommend its execution as a Flagship mission until missions to Enceladus and Uranus have been realized. 

\section{Ganymede and Callisto}\begin{figure}
    \centering
    \includegraphics[width=\textwidth]{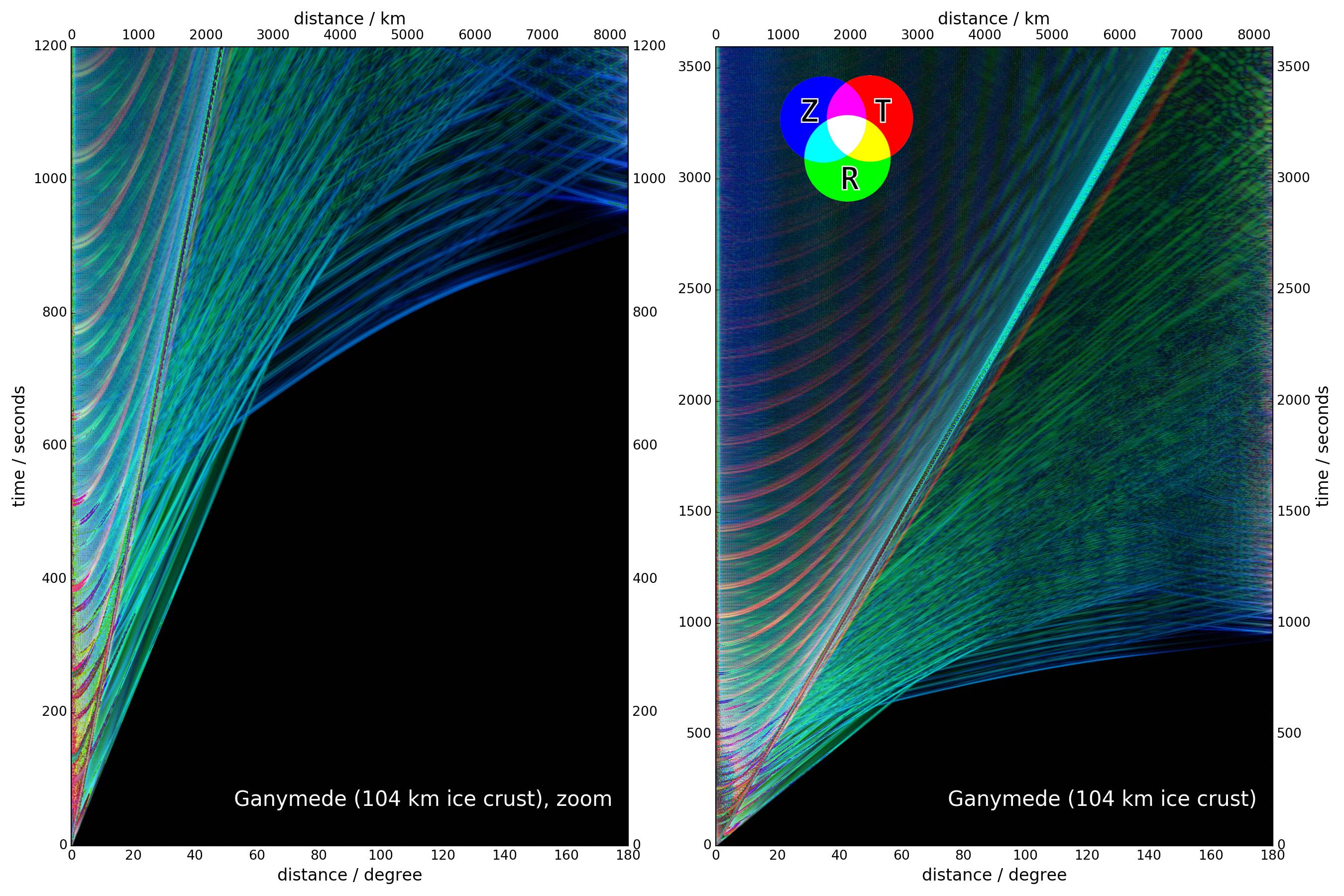}
    \caption{Global seismic wavefield stack for Ganymede, using the interior model of \citet{vance_geophysical_2018}.}
    \label{fig:gs_ganymede}
\end{figure}

\subsection{Potential scientific goals}
Ganymede, Callisto and Titan form the three large ice moons of the solar system and yet, each one has at least one peculiarity that makes it difficult to sum them in one group: Ganymede is the only moon with a current-day magnetic field, Callisto has a very low moment of inertia, suggesting only partial differentiation \citep{nagel_model_2004} and Titan is Titan (see next section). 
For all three planets, the low density implies a ocean deep enough for high-pressure ices to form, inhibiting flow of reductants from the mantle into the ocean  \citep{vance_geophysical_2018}. The presence or absence of these ice layers however depends strongly on the salt content of the ocean, as well as on the poorly known equation of state of salty water at high pressures and low temperatures. Measuring the surface ice thickness could constrain the salt content of the ocean and thus the presence of a high pressure ice layer, even if no direct phases from the high pressure ice can be observed \citep{stahler_seismic_2018}. Seismology could therefore directly address questions of habitability \citep{vance_vital_2018}. In the case of Ganymede, quakes from beyond the core shadow could give insight into the size of a liquid core and the existence of a solid core, as implied by the magnetic field. For Callisto, the question of differentiation of the rocky mantle could be answered much clearer by seismology than by moment of inertia (MoI) measurements from space. As described in chapter 3 \citep{knapmeyer_planetary_2022}, moment of inertia estimates can come with significant errors and lead to long-lived misestimation of core sizes. 
\subsection{Seismicity}
Ganymede shows strong indication of surface tectonics, although its age is difficult to estimate from existing images. Knowledge about current-day tectonics will improve significantly with data from the JUICE mission in the next decade. Callisto's heavily cratered surface indicates that current-day resurfacing and thus tectonic activity is limited.
The higher distance to Jupiter, compared to Europa, means that the tidal energy budget for seismicity is low, but both planets have significant interior heat stored from formation that could drive tectonic deformation.
\subsection{Mission perspectives}
Tidal deformation, will be measured by the Ganymede Laser Altimeter GALA, to estimate the ice layer thickness \citep{enya_ganymede_2022} during the late stage of the JUICE mission, when the spacecraft is orbiting Ganymede. This may be seen as a long period proxy to seismology and likely the closest for some time. Landing on Ganymede or Callisto has been proposed since the Voyager age at least \citep{boain_ganymede_1980} and penetrators have been proposed specifically by European institutions as payloads for Jupiter system flagship missions regularly \citep[e.g.][]{vijendran_penetrator_2010-1}, similar to the Huygens Titan probe on the Cassini mission. Just as regularly, these payloads were cancelled to reduce mission complexity. After JUICE, another Ganymede mission with a lander to either Ganymede or Callisto is not likely to happen before the 2040s. In the meantime, one may watch the 1976 German dystopian movie "Operation Ganymed" starring J\"urgen Prochnow about the difficult return of astronauts from a mission to find life on Jupiter's largest moon.

\section{Titan}
\begin{figure}
    \centering
    \includegraphics[width=\textwidth]{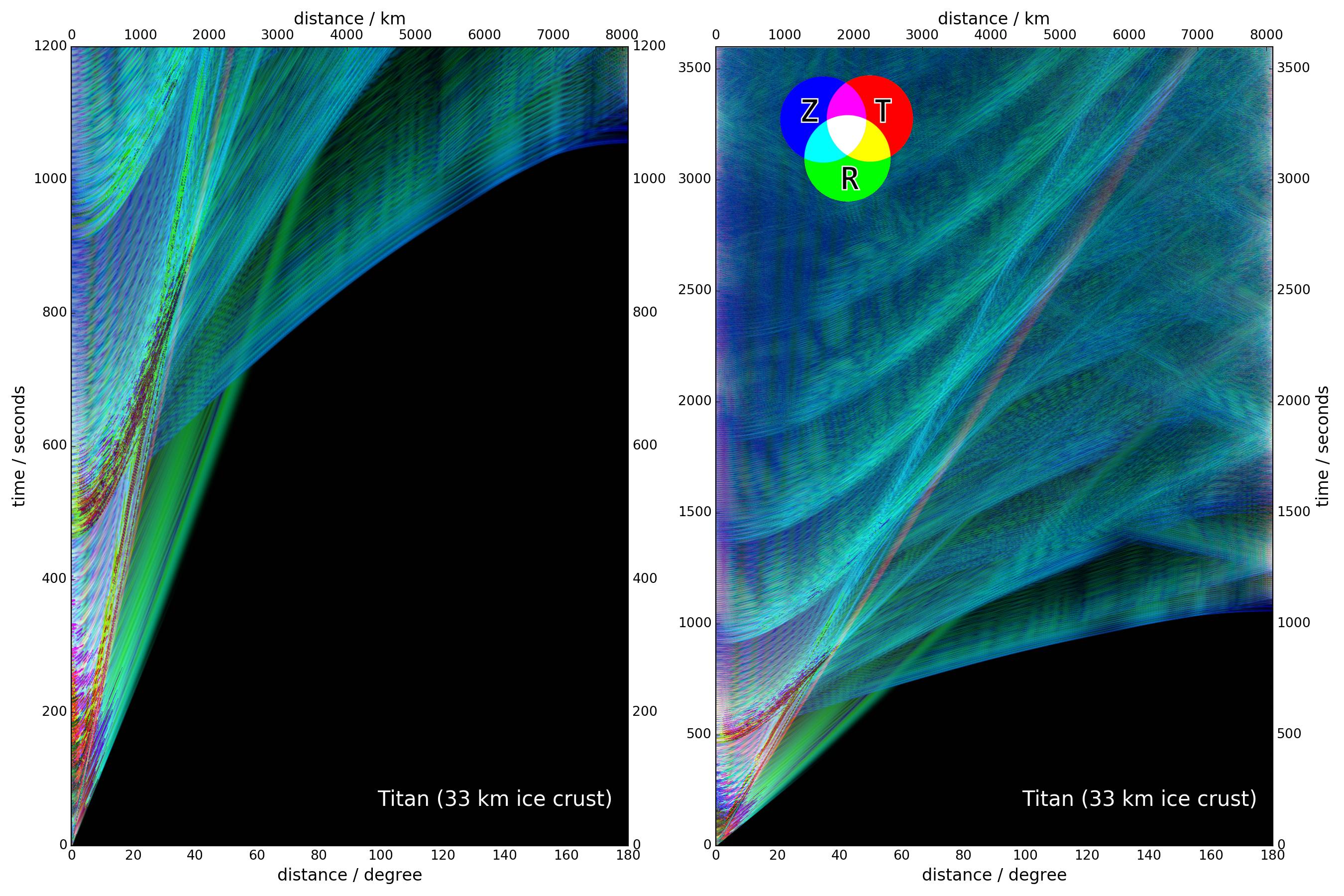}
    \caption{Global seismic wavefield stack for Titan, using an interior model from \citet{vance_geophysical_2018} with a crustal thickness of 33 km.}
    \label{fig:gs_titan}
\end{figure}

\subsection{Potential scientific goals}
Since Titan will be visited by a seismometer in 2034, the science goals of a seismic experiment have been described well already, see e.g. \citet{barnes_science_2021} for an overview. A primary goal is the confirmation of a liquid ocean below the surface via observation of an interface with strong seismic impedance contrast at depth. A peak near 36 Hz in electric field signals measured during the Huygens descent were interpreted as the Schumann resonance of a 55-80 km deep conductor \citep{beghin_analytic_2012}, i.e. the ice-ocean interface. These values are plausible given thermodynamical modelling of the whole ice layer \citep{vance_geophysical_2018}, but far from undisputed. As on other ocean worlds, shear waves will be completely trapped in the ice layer and should lead to well observable reverberations. Compared to other ocean worlds, Titan is likely to harbour methane clathrates (i.e. ice-methane hybrids) near the surface \citep{mousis_methane_2015}, which would have a significant effect on the thermal conductivity and therefore convection of the planet \citep{kalousova_insulating_2020}. These clathrates will have reduced seismic velocities by up to 10\% compared to pure ice \citep{marusiak_methane_2022}, which could be detectable by Rayleigh wave dispersion curves. An open question is the level of viscoelastic attenuation in the ice layer, given the high temperature of the ice below a few 100s of meters of depth. As described in chapter 5 \citep{bagheri_tidal_2022}, ice has a very low quality factor $Q\approx10$ at tidal periods, but the scaling to seismic frequencies is not well constrained.

Due to the existence of large lakes, as well as an atmosphere, Titan might be the only other place in the Solar System in which ocean-generated microseisms can be observed \citep{stahler_seismic_2019}. 
A seismometer deployed north of 60\textdegree~latitude could likely observe the waves created on Kraken mare by a hurricane, enabling remote sensing of the atmosphere.

\subsection{Seismicity}
The tidal deformation of Titan’s ice crust \citep{mitri_hydrocarbon_2007} shown by Cassini gravity measurements is a plausible source of seismic activity. While Titan's orbital period is significantly higher than Europa's or Ganymede's (15.9 vs 3.6/7.2 d), its orbit's high eccentricity could allow tidal forces to drive significant tectonism (see table \ref{tab:tidal}). Whether or not the rocky core of the planet shows significant tectonic activity, is unknown. As on Europa, the science value of detecting quakes from the core would be high, since it seafloor tectonics would enrich the ocean with potential nutrients, but the detection limit is significantly higher than for an iceshell quake.
\subsection{Mission perspectives}
Titan is one of the 2 places in the solar system with a seismic experiment in preparation: The dragonfly lander will be launched in 2027 and deploy a relocatable octocopter on Titan in 2034. The mission will contain a single vertical component seismometer, delivered by JAXA and based on the instrument planned for the Lunar-A mission \citep{mizutani_lunar_1995, shiraishi_present_2008}, plus two or more horizontal geophones. The instrument will not be comparable to the InSight VBB seismometer, but should be able to detect local seismicity.

\section{Enceladus}
\begin{figure}
    \centering
    \includegraphics[width=\textwidth]{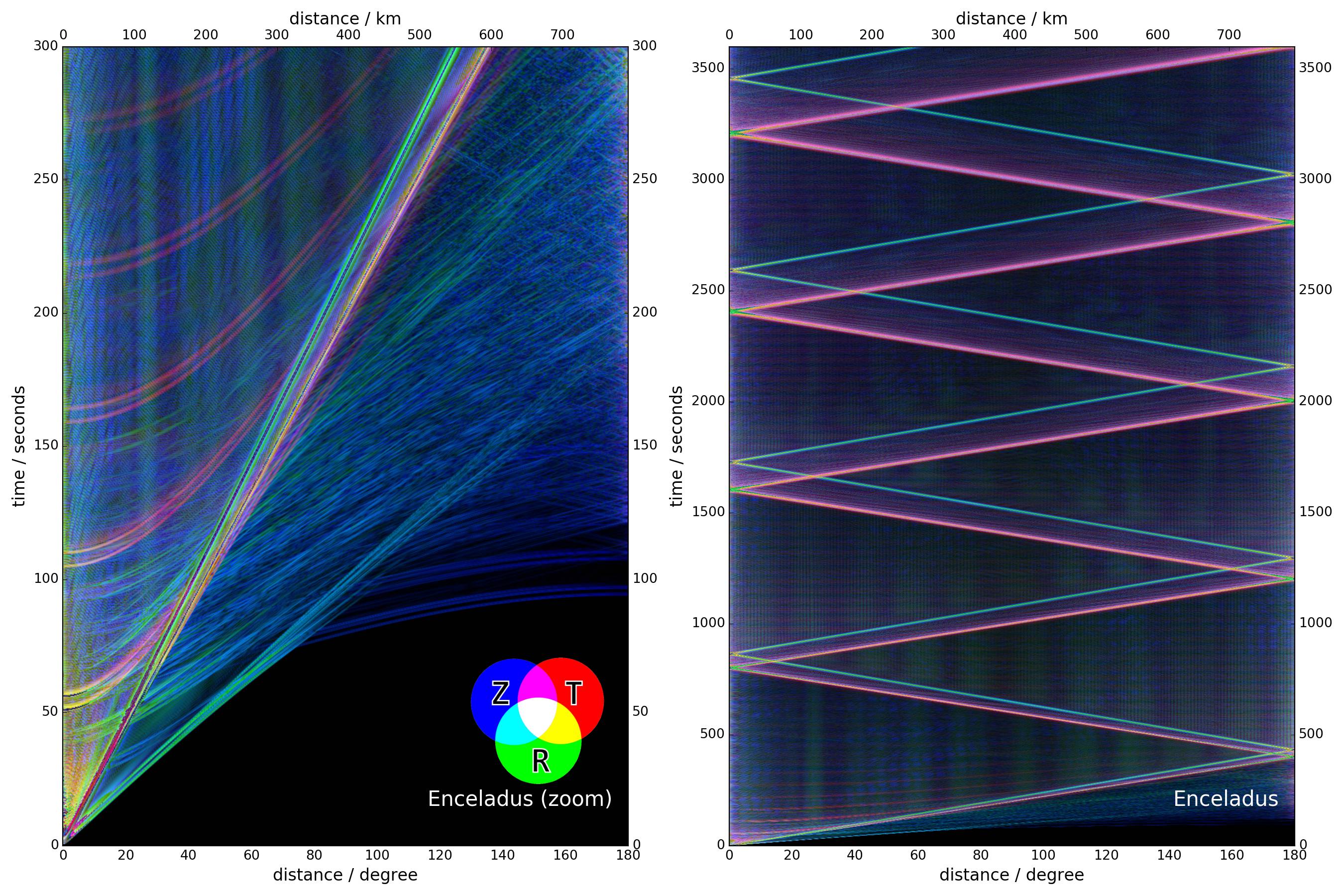}
    \caption{Global seismic wavefield stack for Enceladus, using an interior model from \citet{vance_geophysical_2018}. Note that for modelling reasons, a constant ice thickness of 15 km over the whole planet is assumed, while it gravimetric observations indicate that the ice is significantly thinner near the south pole \citep{porco_cassini_2006}. }
    \label{fig:gs_enceladus}
\end{figure}

\begin{figure}
    \centering
    \includegraphics[width=0.5\textwidth]{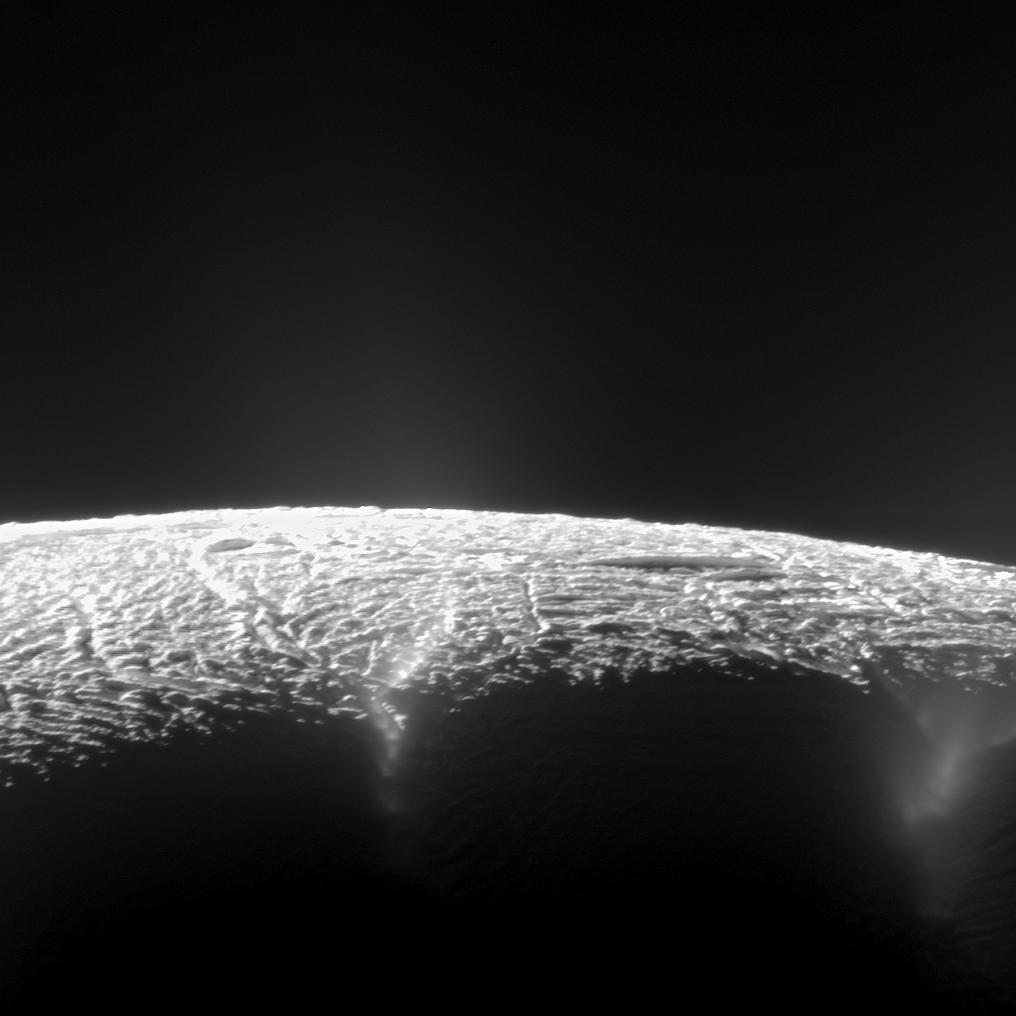}
    \caption{Cassini image of two "tiger stripe" fissures near Enceladus' south pole with water vapor emerging (visible against the dark surface below the terminator). The fissures have been identified as the source of the plumes by \citep{porco_how_2014}. Image by NASA/JPL/Space Science Institute, PIA 17183.}
    \label{fig:enceladus_stripe}
\end{figure}
\subsection{Potential scientific goals}
Enceladus is the only ocean world where water from the subsurface ocean is accessible in situ without drilling through kilometers of ice. A long-lived plume of water vapor and ice has been observed near five distinct fissures (called tiger stripes) near Enceladus' south pole (see fig. \ref{fig:enceladus_stripe}. The Cassini orbiter was able to probe these ejecta directly and found strong indications for a source in the subsurface ocean \citep{teolis_enceladus_2017}. This means that a mission to Enceladus would be directly tasked with determining whether the ocean supports life today or at least has done so in the past \citep{choblet_enceladus_2021}. 

\subsection{Seismicity}
Enceladus has a tidal dissipation larger than Earth's moon, but at a radius of only 252~km. It can therefore be expected that the strain rate of the crust is very significant. The water plume sources near the tiger stripes could produce similar long-duration seismic signals as geysers on Earth.

\subsection{Mission perspectives}
Sampling the plumes could be done from orbit, but the high impact velocities would limit the observability of large molecules (as it has been the case for Cassini's measurements), so the science return of a lander would be significantly higher. The last decadal survey recommended an Enceladus orbiter at low priority, so the design has been updated into a "Orbilander" concept, i.e. an orbiter that is capable of landing on the surface after an extended reconnaissance and remote sensing phase \citep{mackenzie_enceladus_2021}. The current decadal survey \citep{decadal_2023} recommended this mission concept as a flagship mission after finalization of the Uranus orbiter mission, with a launch date after 2037 and landing during favourable communication geometries to Earth in the 2050s.

\section{The Uranus and Neptune system}
\begin{figure}
    \centering
    \includegraphics[width=0.5\textwidth]{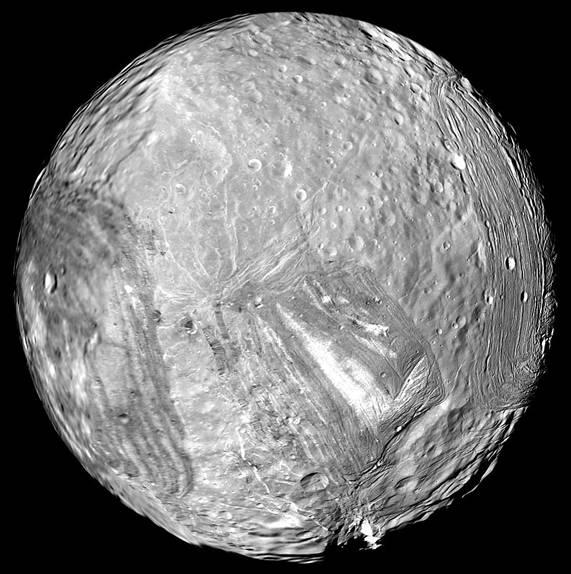}
    \caption{Voyager 2 image of the Uranus moon Miranda. The surface shows linea and TODO. Image by NASA/JPL, PIA 18185.}
    \label{fig:miranda_image}
\end{figure}
Uranus and Neptune are the two ice giants of the solar system and are thereby representing a class of planets that is very common amongst exoplanets discovered so far.  Both of them have a number of planet-like moons that are potential targets for landed missions. However, all of them are known from a few images only, obtained during the flyby of Voyager 2 through the Uranus \citep{stone_voyager_1987} and Neptune systems \citep{stone_voyager_1989} in 1987 and 1989 respectively. Primary science goals would therefore be to research the surface geology, including geomorphology.

\subsection{Potential scientific goals}
Uranus has 4 large moons, Ariel, Umbriel, Titania and Oberon, between 1100 and 1550 km radius, all of which have densities between 1.4–1.7 grams per cubic cm, implying a rock to ice ratio of 2 to 3. A fifth large moon, Miranda, has a diameter of 470 km, but a density of only 1.2 g/cm$^3$, implying an even higher amount of water ice. The surface morphology of Miranda (fig. \ref{fig:miranda_image}) implies that the outer ice layer underwent subduction-like processes in the past, at a similar rate to Europa \citep{hammond_global_2014, stern_stagnant_2018}, which suggests that due to tidal heating, a significant part of this ice was in liquid form at least at one point in the past \citep{nimmo_ocean_2016}. An Uranus orbiter mission will perform dedicated flybys on the large moons and thereby constrain their interiors from gravimetric measurements, similar to what Galileo did in the Jupiter system \citep{sohl_interior_1997}. The question whether the liquid ocean still exists today is one that could be answered by a seismic experiment (see the Europa section for details), but as on Europa and Titan, measuring the seismicity of the planet alone would go a long way in constraining its energy budget.

Neptune has one major moon: Triton, which is comparable in size to the Jovian moon Europa and at an average density of 2.061 g/cm$^3$ widely understood to be covered by several hundred km of frozen or liquid ice \citep{nimmo_ocean_2016}. As on the icy moons of the gas giants, the thickness of the ice and the existence and depth of a potential ocean would be primary science targets for a seismic installation. A unique feature of Triton is its retrograde orbit, which is not found for any other moon of comparable size (the next largest moon, Phoebe with a retrograde orbit around Saturn, has only 0.08 times the mass of Triton), likely the result of capture, possibly due to collision with an original satellite. Similar to Europa, Triton has a relatively young surface age, around 100 Ma \citep{stern_tritons_2000}, as demonstrated by the small number of visible impact craters, implying geological activity over most of the age of the solar system and therefore most likely until today. As on Europa, a relatively thin ice shell opens the possibility of ice convection, potentially even subduction and thus the possibility of large thrust faults with tritonquakes of magnitudes of 5 and larger. 

Of the other neptunian moons, two (Proteus and Nereid) are around 400 km in diameter, but both are irregular in shape, and possibly formed during the capture of Triton \citep{goldreich_neptunes_1989}, when the natural neptunian satellites were cannibalized or deorbited by the newcomer.
\subsection{Seismicity}
No estimations of the level of seismicity of these moons are available up to now.
\subsection{Mission perspectives}
The Decadal Survey 2023-2033 \citep{decadal_2023} recommended a flagship mission to one of the ice giants, and preferred a Cassini-like orbiter in the Uranus system over one to Neptune, on grounds of feasibility. Specifically, launch opportunities arise in 2031 and 2032 for a 13 year cruise without the need of inner solar system gravity assists. A Neptune mission would require significant technical development, specifically rely on the SLS rocket, while planning for a Uranus orbiter could start immediately. ESA's "Voyage 2050" report \citep{tacconi_esa_2021} has expressed interest in contributing an atmospheric entry probe or moon lander to a NASA flagship mission to an ice giant, similar to the Cassini/Huygens contribution, where ESA delivered the Huygens Titan lander. The next years will show whether this contribution will materialize and whether it will be in form of a Uranus entry probe or a lander on Miranda or another large moon.

A mission to the Neptune system is therefore very unlikely to be even considered before the 2040s. At this time, there would likely be a push for such a mission to contain at least a short-lived lander probe to Triton.

Due to the distance to the sun, solar power is not feasible for sustained operation and as for the Europa Lander, such a mission would have to rely on high performance batteries or RTGs.


%
%
\section{Interstellar objects}
\subsection{Potential scientific goals}
The first discovery of an interstellar object is less than five years old, when 1I/2017 U1 (‘Oumuamua) was observed on 2017 October 19 by the Pan-STARRS1 telescope system and confirmed to be travelling on a hyperbolic orbit, i.e. not been bound by the sun's gravity \citep{meech_brief_2017}. The limited amount of observations that were possible in the short time window of observation confirmed a rocky, red surface without any trace of degassing, as it would be observed for a comet from the Solar System's Oort cloud \citep{bannister_natural_2019}. Surprisingly, a second interstellar object was found only 2 years later, this time before aphelion, so that its dynamical behaviour during approximation of the sun could be observed and the object 2I/Borisov could be confirmed to be an ice-rich comet \citep{bodewits_carbon_2020, guzik_initial_2020}.

The investigation of interstellar objects is obviously interesting. Interstellar objects entering the solar system on hyperbolic trajectories are the only solid matter from outside the solar system that will be accessible to in situ characterisation in the foreseeable future. A lander mission would carry simple instruments to obtain a chemical composition of the top layers. The rigidity of the object however is difficult to assess from the outside, but the distinction on whether the object is a homogeneous body, an ice-rock mixture or even a "rubble ball", a very weakly consolidated object, is of high interest to constrain the source context. The SESAME/CASSE seismic experiment on the Philae mission to comet 67P/Churyumov-Gerasimenko showed a mechanically extremely weak core below a harder surface layer \citep{knapmeyer_structure_2018}. The latter is likely the result of previous encounters of the comet to the sun \citep{groussin_thermal_2019}. Whether similar layering exists for true interstellar objects would be quite interesting. As the contact of the TAG sampler on the OSIRIS-REx mission to Bennu showed, even asteroids can have surprisingly low rigidity in its uppermost layers \citep{berry_contact_2022}.
\subsection{Seismicity}
Interstellar objects of the size encountered so far are unlikely to sustain significant tectonic activity themselves. A possible mechanism for seismic sources is thermal stress induced by the approximation of the sun or cooling while leaving the solar system. Another option would be the combination of a slow impactor mission with a landed seismometer or some kind of repeatable source on the lander, like the hammering devices used on Philae and InSight, or the mortars used on Apollo 14 and 16.

\subsection{Mission perspectives}
The observation of two interstellar objects in relatively short time has triggered significant interest in developing a mission for in situ exploration of objects to be detected in the near future \citep{hein_project_2019, seligman_feasibility_2018, castillo-rogez_approach_2019}. Proposals exist for missions that are prepared and wait in storage for the discovery of a suitable object, either on Earth, or the sun-earth L2 point. Even though fly-by missions are significantly easier, a lander using solar-electric propulsion could be feasible \citep{hein_interstellar_2022}. A landed mission would bear similarity to Philae on the ESA Rosetta mission and could do seismic or acoustic investigations to constrain subsurface properties either during landing \cite{ biele_landings_2015} or by listening to seismic waves excited during operation of a drill or similar instrument \citep{knapmeyer_sesamecasse_2016, knapmeyer_structure_2018}.
The intercept point to any interstellar object would likely be outside of Jupiter's orbit and therefore solar energy is not an option for a small lander. The lifetime of a surface mission would therefore be limited by battery capacity and seismic experiments would be coordinated with sampling or impactor operations. The primary mission of Philae on comet 67P, sustained by the primary battery only, for example, was about 68 hours.

\section{Lessons}
\subsection{Scattering}
Terrestrial seismology builds on clearly separated arrivals of ground motion, called "phases". Only the fact that seismic waves travel mostly unperturbed through the earth and thus arrive in short pulses made it possible to disentangle the plethora of signals being reflected and converted at various layers and interfaces inside the planet. A necessary conditions for such clean phases is that the length scale of heterogeneities inside the planet is larger than the wave lengths involved \citep{aki_quantitative_2002}. The first seismograms observed on the moon showed that this condition is not fulfilled there and that instead, the lunar crust scatters all seismic phases beyond recognition \citep{blanchette-guertin_investigation_2012}. The classical explanation for this is a high impact rate due to the lack of an atmosphere and a lack of mechanisms to heal small cracks, due to the absence of fluids, combined with very low intrinsic attenuation, again due to extremely low water content. Preliminary analyses of Martian seismic data showed surprisingly high amount of scattering \citep{menina_energy_2021, karakostas_scattering_2021}, at least close to the InSight landing site. Since this scattering seems limited to the uppermost kilometers, investigations of "clean" mantle phases were possible nevertheless \citep{duran_seismology_2022}. 

It is likely however, that at least all airless rocky planets (Mercury) will be more similar to the Moon in terms of problematic scattering. Strongly cratered icy worlds like Ceres or Callisto might be similar, while the tidal heating in Europa, Titan and Enceladus may lead to ductile deformation of the crust and healing of heterogeneities.

This scattering will affect single station, three component seismic measurements. It is possible that gradiometric measurements, e.g. measurements of rotational motion \citep{bernauer_rotation_2021} or distributed measurements at multiple locations \citep{walter_distributed_2020} could allow to detect coherent wavefronts even in presence of strong scattering. However, sensitivity of such sensors is typically too low for planetary applications still \citep{bernauer_exploring_2020}, even though the field is rapidly progressing. 

\subsection{Timing} 
The Sesame Casse experiment on the ESA Rosetta mission was the first seismic experiment that used vibrations from a mechanical hammer as source in 2014. An issue encountered in the analysis, was the limited coordination between instruments. A dedicated trigger line from the MUPUS hammer to the SESAME recording system was considered too complex in the preparation phase. Instead, the two instruments exchanged messages in a common part of the onboard computer memory to coordinate hammering and recording, and an on-ground assessment of the seven involved clocks and their individual drift rates was carried out during the evaluation \cite{knapmeyer_sesamecasse_2016}. This made initial experimentation much more difficult \cite{knapmeyer_structure_2018}. The Rosetta mission was launched in 2004, the problem was encountered in 2014, but surprisingly, the same problem occurred again during the InSight HP3 seismic experiment, launched in 2018, when the seismometer SEIS was listening for seismic waves produced by the hammer of the HP$^3$ heat flow probe \citep{spohn_insight_2021, sollberger_reconstruction_2020}. The lack of a joint time signal between the two instruments HP$^3$ and SEIS meant that a convoluted process was necessary to infer the exact time of each hammer blow from the seismic signal itself, significantly increasing the uncertainty of the observation. Since the distance between any hammering or drilling instrument and a seismic sensor is unlikely to be more than a few meters and seismic velocities $>500$ m/s are expected, the precision of the joint timing source needs to be $<100 \mu$s, which is below the typical resolution of spacecraft bus clocks. 
\subsection{Bandwidth}
InSight was able to transmit 6 seismic channels of 20 sps each over much of the mission due to availability of orbiters with relay capacity (Odyssey, Mars Reconnaissance Orbiter and the ESA Trace Gas Orbiter). This far exceeded the pre-mission planning, where a single 10 sps vertical channel and the 3 native channels of the seismometer were foreseen. In retrospect, one event type, the "super high frequency events", likely thermal cracking near the lander, would not have been detected with the original configuration \citep{dahmen_super_2020}, while another type, the very high frequency events, would have been much more difficult to spot, given that their energy is mainly on the horizontal channels above 2 Hz.
It was specifically foreseen to retrieve higher bandwidth data of specific events by manual requests. Such events were either to be detected in the low bandwidth streams or in one specific "ESTA-SP" channel, that contained the integrated signal energy in a narrow frequency band above 10 Hz. This process worked successfully overall, with the caveat that the ESTA-SP channel was polluted too much by glitches and wind-related transient signals to be of much use overall.

Any future seismic mission (save for a lunar one) will likely be dramatically more constrained in terms of bandwidth, so that only a subset of data can be transmitted. At the same time, low sampling rates risk omitting interesting signals of local events. Classic, lossless seismic compression \citep{ahern_seed_2012} reduces the amount of data by 30-50\%, which is by far not enough. 

\begin{figure}
    \centering
    \includegraphics[width=0.8\textwidth]{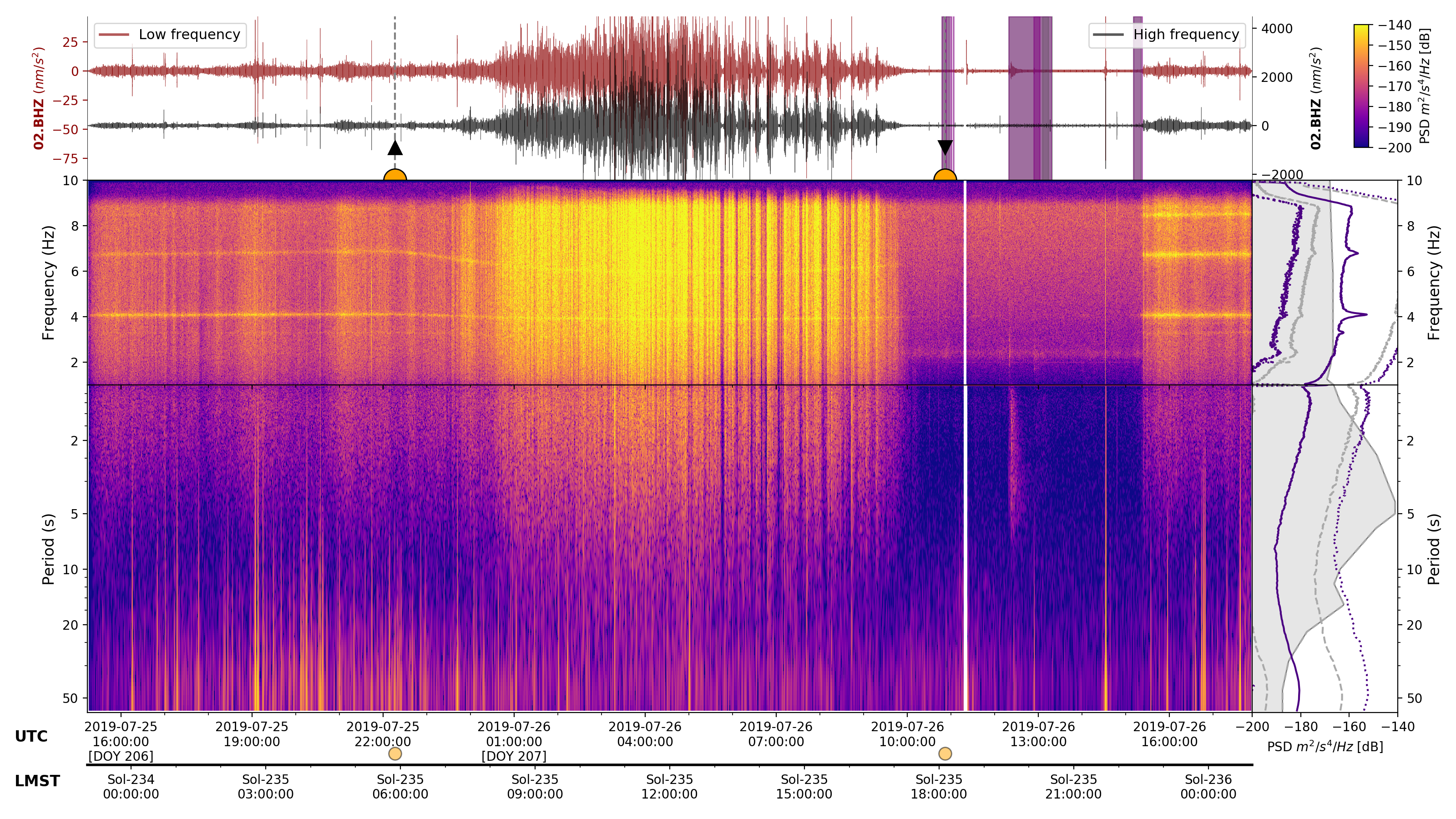}
    \includegraphics[width=0.73\textwidth]{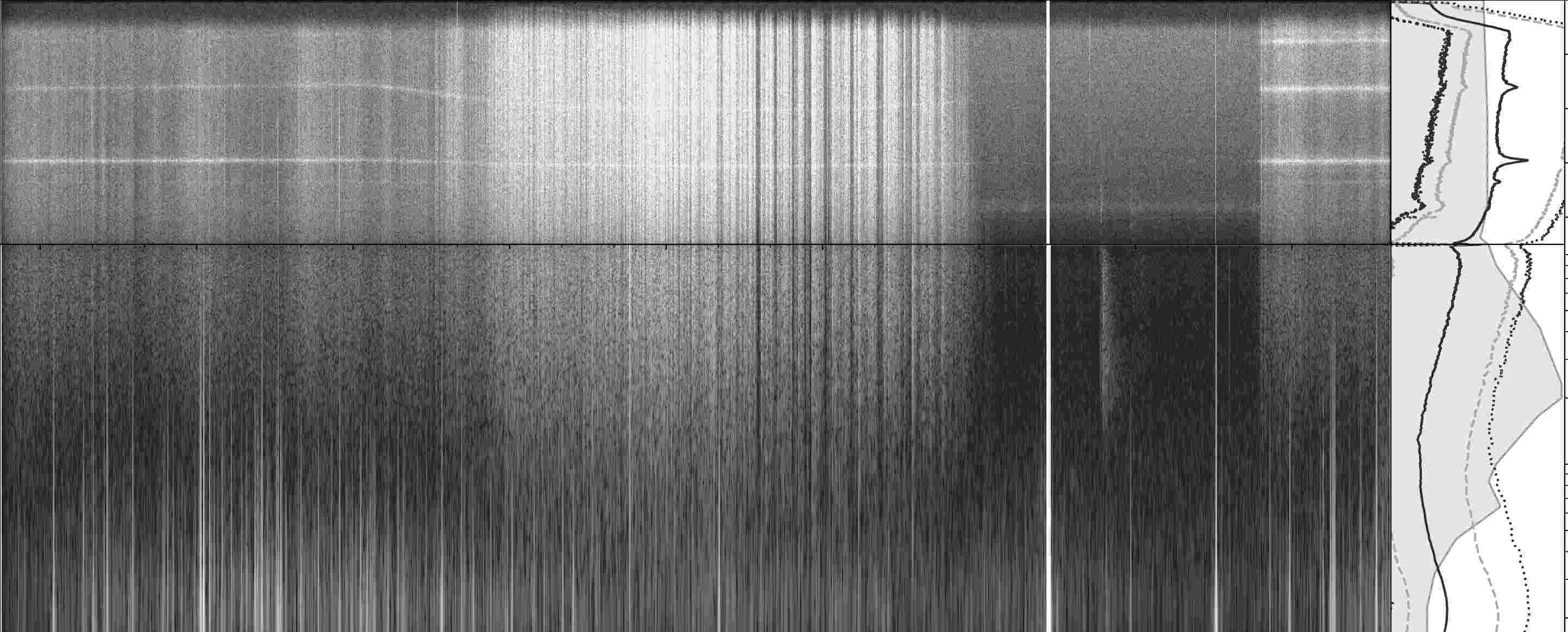}
    \caption{top: Spectrogram of one Sol of seismic data, plot in the style of      
             \citep{giardini_seismicity_2020}, fig. 1a. 3 marsquakes can be identified easily.
             bottom: Same plot, subjected to level 10\% lossy JPEG compression. The size of this figure is 108 kByte, i.e. 2 \% of the original seismic data, yet the marsquake time windows can still be identified.}
    \label{fig:spec_compression}
\end{figure}

Another possibility would be advanced pre-processing of data onboard. Figure \ref{fig:spec_compression}a shows a spectrogram of 26 hours of vertical component VBB data of InSight, as described by \citep{giardini_seismicity_2020}. Over the course of the day, the wind noise increased dramatically over noon, masking all potential marsquakes. In the evening, S0235b, the highest SNR event of the first Martian year, can be seen. This spectrogram is equivalent to $24.7 ~ 3600 ~ 20 ~ 24 = 41.4$~MBit or 5.2 MByte of seismic data. As described in \citep{clinton_marsquake_2021}, operationally, all significant marsquakes can be easily detected in one such spectrogram. Figure \ref{fig:spec_compression}b shows the same spectrogram as JPEG graphic, stored with high compression. The 3 marsquakes can still be detected in this graphic, yet the file size of this graphic is 108 kByte, i.e. 2\% of the original seismic data. From this small file, interesting time windows could easily be identified for transfer of data at full bandwidth. Whether spectrograms or similar, wavelet based time-frequency representations are the optimal way of producing low bandwidth daily overview files needs to be investigated and weighted against the computational power available onboard. 

For certain mission concepts in the outer solar system, e.g. the Europa Lander \citep{hand_report_2017}, the short mission duration will require the lander to exercise its scientific campaign mostly autonomously without "Earth in the loop". This may mean that seismic events need to be identified autonomously by the lander itself. Given the phenomenology of non-seismic events, even in terrestrial planets \citep{ceylan_companion_2021, dahmen_resonances_2021-1, stahler_geophysical_2020}, let alone on icy ocean worlds, this process is non-trivial and will require significant theoretical work over the next years. Given the surprises we saw with both moonquakes and marsquakes, it is unlikely that a fully autonomous processing can find all kinds of events without prior knowledge of typical examples.


\section{Conclusion}
The InSight mission to Mars has demonstrated that even a single, two year seismometer mission is enough for a first global determination of the deep interior of a planet. It did confirm geodetic observations from orbit concerning the core radius, as well as imaging-based inferences of current day tectonics in Cerberus Fossae. Furthermore, it allowed to constrain previously inaccessible parameters, such as the crustal thickness or the rate of quakes. Finally, unexpected discoveries were made, such as the high amount of scattering in the shallow crust. The seismic observations, like travel times also provide a quantitative dataset to test future interior models of Mars against for the coming decades. 

A similar dataset would be highly valuable for any of the other planets and moons of the solar system. Yet, given the difficulty and cost of landing, it needs to be weighted against orbital data, geodetic, radar or image based, which is easier to obtain. From a purely subjective perspective, 4 questions stand out that can be only addressed by seismology.

\begin{enumerate}
    \item The tectonic context of samples on Ceres. A Ceres sample return mission in Occator crater would rely heavily on knowledge of the age and deposition mechanism of these briny samples. Even a low-sensitivity seismometer could pick up small quakes near the landing site to determine whether signals related to active cryovolcanism exist and suggest a young age of the deposits.
    \item The ice thickness and ocean depth of Europa and Titan. The Europa Clipper radar will have a good shot at finding a reflection from the ice-ocean interface, but the success of this measurement is subject to the poorly constrained electromagnetic absorption in the icy crust. Yet, the crustal thickness and ocean depth are the strongest quantitative constraints on composition of the ocean and thus its habitability. A seismic experiment could give an independent constraint. Hopefully, the Dragonfly seismometer will perform good enough to prove this concept.
    \item The formation history and tidal heating of the large moons. The four Galilean moons differ so strongly in their interior profiles (as constrained from geodetic data) that it is difficult to judge whether our formation models work on them. Given that tidally heated planets and moons in resonance might be significantly more common in other planetary systems (think of Trappist-1), constraining even only the interior of Ganymede or Io seismically would go a long way in calibrating the models we apply to exoplanets.
    \item The meteoroid impact rate throughout the solar system. At the time of writing, there are no confirmed detections of meteoroid impacts by InSight, even though candidate signals exist and are reviewed. The rate of crater formation is assumed to be known when determining the age of planetary surfaces, yet the rate of new impacts cannot be constrained well, due to the strong effect of target material on observability of fresh craters. A seismic observatory on a planet can constrain this rate directly, which also has a direct implication for estimating the density of small bodies in the solar system and the resulting hazard for Earth and interplanetary spaceflight.
\end{enumerate}
The experience from InSight will make it possible to build smaller, low-power seismometers, which can be added to any future landed mission at a low cost in terms of complexity and energy.
\printbibliography
\end{document}